\documentclass[aps,prx,superscriptaddress,showpacs,twocolumn]{revtex4-1}

\usepackage{graphicx}
\usepackage{dcolumn}
\usepackage{bm}
\usepackage{color}
\usepackage{xcolor}
\usepackage{amsmath}
\usepackage{amssymb}
\usepackage{amsmath}
\usepackage{float}
\usepackage{natmove}
\usepackage{multirow}
\usepackage{setspace}
\usepackage{array}
\usepackage{booktabs}
\usepackage{rotating}
\usepackage{filecontents}
\usepackage{xr-hyper}
\makeatletter
\newcommand*{\addFileDependency}[1]{
  \typeout{(#1)}
  \@addtofilelist{#1}
  \IfFileExists{#1}{}{\typeout{No file #1.}}
}
\makeatother

\newcommand*{\myexternaldocument}[1]{%
    \externaldocument{#1}%
    \addFileDependency{#1.tex}%
    \addFileDependency{#1.aux}%
}
\usepackage[colorlinks=true,linkcolor=blue,citecolor=blue,urlcolor=blue]{hyperref}
\usepackage{physics}
\usepackage{mathrsfs}
\usepackage{tabularx}
\usepackage{layouts}
\usepackage{ulem}

\newcommand{\suppinfo}{Supplemental Material~\cite{supp-info}}
\newcolumntype{C}{>{\centering\arraybackslash}X}

\definecolor{tangerine}{rgb}{0.944,0.522,0}
\definecolor{brown}{rgb}{0.633,0.156,0.156}
\definecolor{lime}{rgb}{0.5,1.0,0.0313}
\definecolor{limedark}{rgb}{0.333, 0.666, 0.020}
\definecolor{applegreen}{rgb}{0.55, 0.71, 0.0}
\definecolor{green1}{rgb}{0.0, 0.5, 0.0}
\definecolor{green2}{rgb}{0.25, 0.5, 0.016}
\definecolor{BluBondi}{rgb}{0.00,0.58,0.71}
\definecolor{myred}{rgb}{0.784, 0.063, 0.180}  
\definecolor{mygreen}{rgb}{0.478,0.604,0.004}  
\definecolor{myblue}{rgb}{0.059,0.298,0.506}   

\newcommand{\editor}[2]{%
  \expandafter\newcommand\csname #1note\endcsname[1]{%
    \textcolor{#2}{(\textbf{#1:} ##1)}}%
  \expandafter\newcommand\csname #1\endcsname[1]{%
    \textcolor{#2}{##1}}%
  \expandafter\newcommand\csname #1cancel\endcsname[1]{%
    \textcolor{#2}{\sout{##1}}}%
  \expandafter\newcommand\csname #1change\endcsname[2]{%
    \textcolor{#2}{\sout{##1} ##2}}%
  \newenvironment{#1text}{\color{#2}}{\color{black}}
}

\editor{AF}{orange}
\editor{NM}{green}
\editor{TC}{red}
\editor{note}{BluBondi}

\myexternaldocument{supplemental_material}

\begin{document}
\title{A unified Green's function approach for spectral and thermodynamic properties from algorithmic inversion of dynamical potentials}

\author{Tommaso \surname{Chiarotti}}
\email[corresponding author: ]{tommaso.chiarotti@epfl.ch}
\affiliation{Theory and Simulations of Materials (THEOS)
             and National Centre for Computational Design and Discovery of Novel Materials (MARVEL),
           \'Ecole Polytechnique F\'ed\'erale de Lausanne, 1015 Lausanne, Switzerland}
\author{Nicola \surname{Marzari}}
\affiliation{Theory and Simulations of Materials (THEOS)
             and National Centre for Computational Design and Discovery of Novel Materials (MARVEL),
           \'Ecole Polytechnique F\'ed\'erale de Lausanne, 1015 Lausanne, Switzerland}
\author{Andrea \surname{Ferretti}}
\affiliation{Centro S3, CNR--Istituto Nanoscienze, 41125 Modena, Italy}

%
%
\date{\today}

%
%
%
\begin{abstract}

Dynamical potentials appear in many advanced electronic-structure methods, including self-energies from many-body perturbation theory, dynamical mean-field theory, electronic-transport formulations, and many embedding approaches. 
Here, we propose a novel treatment for the frequency dependence, introducing an algorithmic inversion method that can be applied to dynamical potentials expanded as sum over poles. This approach allows for an exact solution of Dyson-like equations at all frequencies via a mapping to a matrix diagonalization, and provides simultaneously frequency-dependent (spectral) and frequency-integrated (thermodynamic) properties of the Dyson-inverted propagators.
The transformation to a sum over poles is performed introducing $n$-th order generalized Lorentzians as an improved basis set to represent the spectral function of a propagator, and using analytic expressions to recover the sum-over-poles form.
Numerical results for the homogeneous electron gas at the $G_0W_0$ level are provided to argue for the accuracy and efficiency of such unified approach.

\end{abstract}

%
%
\maketitle
%

\section{Introduction}
\label{sec:intro}
Electronic-structure calculations have been and remain a powerful and ever expanding field of research to understand and predict materials properties~\cite{marzari_electronic-structure_2021}. 
The development of methods, algorithms, and hardware brings in continuous progress, allowing for computational materials discovery~\cite{hafner_toward_2006,curtarolo_high-throughput_2013,mounet_two-dimensional_2018}, accurate comparison with experiments~\cite{zhou_unraveling_2020,reining_gw_2018}, and even hybrid quantum-computation algorithms~\cite{ma_quantum_2020,ma_quantum_2021}.

Due to the interaction between the electrons in a system, solving the many-body quantum problem is often at the core of many approaches. Focusing on condensed-matter systems, density-functional theory (DFT) has been one of the most used and successful methods so far~\cite{van_noorden_top_2014}.
The possibility to map exactly the ground-state solution of the $N$-body problem to the minimization of a density functional for the energy~\cite{hohenberg_inhomogeneous_1964} offers great computational simplifications, allowing the accurate computation of ground-state quantities for most materials.
Although mathematically well-defined~\cite{levy_universal_1979} and computationally inexpensive, it remains challenging to improve the approximate functionals~\cite{perdew_generalized_1996,sun_strongly_2015}  --- often resulting in incorrect predictions for complex or strongly-correlated systems~\cite{cohen_insights_2008} --- or to address spectroscopic properties~\cite{perdew_understanding_2017,nguyen_koopmans-compliant_2018}.

Dynamical (i.e. frequency-dependent) theories like many-body perturbation theory, dynamical mean-field theory, and in general embedding theories offer the flexibility to overcome these limitations of DFT. 
While the type of embedding differs in different approaches, a common element is the appearance of dynamical potentials.
As an example, many-body perturbation theory (MBPT) reduces the multi-particle electronic degrees of freedom to one via frequency embedding~\cite{Martin-Reining-Ceperley2016book}. 
Dynamical mean-field theory (DMFT) couples a real-space impurity with the rest of the system, requiring self-consistency between the two self-energies acting on the impurity and on the bath~\cite{georges_dynamical_1996}.
Self-energy-embedding theory (SEET) calculates exactly the frequency-dependent self-energy of strongly-correlated manifolds in solids, and applies it to the remaining weakly interacting orbitals~\cite{kananenka_systematically_2015,lan_generalized_2017}.
Coherent electronic-transport theories use a Green's function embedding to calculate the electronic conductance of e.g. a conductor between two semi-infinite leads, coupling the three systems dynamically~\cite{calzolari_ab_2004,ferretti_maximally_2007}.
Clearly, handling properly frequency-dependent potentials is of central interest in the field.

Using here MBPT as a paradigmatic example, we highlight that the difficulty in treating dynamical quantities has often led to different methodological approaches when calculating spectral or thermodynamic quantities (such as energies, number of particles, chemical potentials). Real-axis calculations are commonly performed to compute the frequency-dependent spectral properties~\cite{hedin_transition_1998,damascelli_angle-resolved_2003,golze_gw_2019}, while the frequency-integrated thermodynamic properties are typically calculated using an imaginary-axis formalism~\cite{garcia-gonzalez_self-consistent_2001,schindlmayr_diagrammatic_2001,dahlen_variational_2004,dahlen_variational_2006,pavlyukh_dynamically_2020,kutepov_electronic_2016}. 
In a series of papers~\cite{von_barth_self-consistent_1996,holm_self-consistent_1997,holm_fully_1998,holm_total_2000} von Barth and coworkers have proposed a formalism partially able to tackle spectra and thermodynamics together for the homogeneous electron gas~\cite{Giuliani-Vignale2005book}, by modelling the spectral function in frequency-momentum space using Gaussians with $k$-parametrized centers (quasi-particle energies), broadening (weights), and satellites. 
Due to its model nature, the approach does not easily offer the flexibility to target realistic systems and in general extend to embedding problems.

Here we introduce a novel approach, termed algorithmic-inversion method, applied on sum-over-pole expansions (AIM-SOP), to address the simultaneous calculation of accurate spectral and thermodynamic quantities. Within AIM-SOP, dynamical (frequency-dependent) self-energies are expanded on sum over poles, and the exact solution --- at all frequencies --- of the Dyson equation is found via a matrix diagonalization.
The transformation of a frequency-dependent propagator into a SOP via a representation of its spectral function on a target basis set is greatly improved with the introduction of $n$-th order generalized Lorentzians as a basis with improved decay properties. 
The SOP form allows one to compute analytically convolutions and moments of propagators for the calculation of spectral, and thermodynamic properties.
Owing to the fulfillment of all sum rules implied by the Dyson equation, we show that the AIM-SOP method becomes essential to have accurate frequency-integrated quantities in a real-axis (thus, spectral oriented) formalism. 
As a case study, we consider the paradigmatic case of the homogeneous electron gas (HEG), for $r_s$ from $1$ to $10$, treated at the $G_0W_0$ level~\cite{lundqvist_single-particle_1967,lundqvist_single_1967,lundqvist_single-particle_1968}.

The paper is organized as follows: In Sec.~\ref{sec:method_SOP} we introduce the AIM-SOP approach, discussing its main goal and the SOP form for propagators and self-energies.
In Sec.~\ref{sec:method_representation} we provide an overview  of the connection between a propagator and its spectral function, first for a continuum and then extending it to treat spectral functions represented on discrete basis sets, as will be used in this work. 
Then, we consider different basis sets to represent the spectral function and obtain a SOP representation introducing $n$-th order Lorentzians. 
In Sec.~\ref{sec:method_transform} we provide the numerical procedure to transform a propagator sampled on a grid to a SOP representation, and viceversa. In Sec.~\ref{sec:method_analytical} we introduce several useful expressions when dealing with propagators on SOP, such as analytic convolutions and moments, and in Sec.~\ref{sec:validation} we show with a numerical example the representation on SOP for a test propagator. 
Finally, in Sec.~\ref{sec:Dyson_SOP} we present the algorithmic-inversion method on sum over poles to obtain exact solutions on SOP of any Dyson-like equation. We first provide a mathematical proof for the case of a self-energy on SOP, then we discuss the case of the polarizability inversion, providing a numerical example as proof-of-concept for the procedure.
In Sec.~\ref{sec:method_HEG} we discuss the application of the method to the test case of the homogeneous electron gas. In Sec.~\ref{sec:results} we discuss the results obtained applying AIM-SOP to the homogeneous electron gas at the $G_0W_0$ level, first discussing the $r_s=4$ case in detail and then presenting results for $r_s$ from $1$ to $10$.
Finally, in Sec.~\ref{sec:conclusion} we draw the conclusions for the paper. Technical aspects of the method are further presented in the Appendices.

\section{Method: AIM-SOP for dynamical potentials}
\label{sec:method_SOP}
%
In this Section we introduce the algorithmic-inversion method to treat dynamical (frequency-dependent) potentials. The crucial goal for AIM-SOP is to solve exactly and at all frequencies Dyson-like equations for dynamical potential expressed as sum over poles.
For this purpose we express frequency-dependent propagators and self-energies (or, say, polarizabilities or screened Coulomb interactions) in a SOP form:
\begin{equation}
    G(\omega)=A_0 + \sum^N_{i=1} \frac{A_i}{\omega-z_i},
    \label{eq:SOP}
\end{equation}
where the constant term $A_0$ may be present for self-energies and potentials.
Generally, we consider here having complex residues $A_i$ and poles $z_i=\epsilon_i+i\delta_i$ ($\epsilon_i,\delta_i\in \mathbb{R}$).
In order to provide the correct analytical structure respecting time-ordering, $\delta_i\gtrless0$ when $\epsilon_i\lessgtr \mu$, where $\mu$ is the effective chemical potential of the propagator (for a Green's function  $\mu$ is the Fermi energy of the system, for a polarizability or a screened potential $\mu=0$). 

Throughout this work we will use as case of study the homogeneous electron gas (HEG) also in view of to the extensive algorithmic and numerical results in the literature. In the HEG, due to translational symmetry, the two-point operators (including Green's functions, self-energies, polarizabilities) are diagonal on the plane-wave basis, but the AIM-SOP can be generalized to non-homogeneous systems, as will be discussed in future work.

\subsection{Spectral representations}
\label{sec:method_representation}
%
Following Ref.~\cite{holm_self-consistent_1997}, we consider the spectral representation of a propagator (here the Green's function for simplicity), where $G$ is expressed in terms of its spectral function $A$,
\begin{eqnarray}
    G(\omega)&=&\int_{\mathcal{C}} \frac{A(\omega')}{\omega-\omega'} \, d\omega'
    \label{eq:TOHT}
\end{eqnarray}
by performing a
time-ordered Hilbert transform (TOHT), where $\mathcal{C}$ is a time-ordered contour which is shifted above/below the real axis for $\omega' \lessgtr\mu$, 
and where the shift is sent to zero after the integral is computed.
%
Accordingly, the inverse relation to go from $G$ to $A$ is given by:
\begin{eqnarray}
  \label{eq:G_to_A}
  A(\omega)&=&\frac{1}{2\pi i}\left[ G(\omega) -G^\dagger(\omega) \right]\,\text{sgn}(\mu-\omega) \\
  &=& \frac{1}{\pi} \text{Im}G(\omega)\,\text{sgn}(\mu-\omega),
  \nonumber
\end{eqnarray}
the last expression being valid for a scalar Green's function, as is the case for the HEG.
%
Representing the spectral function on a (finite) basis set $\{b_j(\omega)\}$,
\begin{equation}
    A(\omega)=\sum_j a_j b_j(\omega)\, \text{sgn}(\mu-\epsilon_j) = 
    \sum_j a_j 
    \abs{b_j(\omega)},
    \label{eq:A_representation}
\end{equation}
with $b_j(\omega)$ centred on $\epsilon_j$ and positive (negative) for $\epsilon_j\lessgtr\mu$, respectively, and $a_j>0$, we also induce a representation of $G$. This is achieved by introducing a discrete time-ordered Hilbert transform (D-TOHT) as
\begin{eqnarray}
    G(\omega)&=& \sum_{j} a_j \int \frac{\abs{b_j(\omega')}}{\omega-\omega'-i0^+ \ \text{sgn}(\epsilon_j)} \, d\omega',
    \label{eq:TOHTdiscrete}
\end{eqnarray}
where the sign chosen for $b_j$ in Eq.~\eqref{eq:A_representation} gives by construction the time-ordered analytic structure of the Green's function.
In the case of all $\delta_j\to0$ with the number of $b_j$ becoming infinite (continuum representation limit), Eq.~\eqref{eq:TOHTdiscrete} becomes the standard TOHT of Eq.~\eqref{eq:TOHT} (with $\mathcal{C}$ shifted by $\pm i0^+$). 


A natural choice is to use a basis of Lorentzian functions centered at different frequencies $\epsilon_j$, according to:
\begin{equation}
    |b_j(\omega)|=\mathcal{L}_{\delta_j}(\omega-\epsilon_j) = \frac{1}{\pi} \frac{\abs{\delta_j}}{(\omega-\epsilon_j)^2+\delta_j^2},
    \label{eq:1lor_basis}
\end{equation}
for which the D-TOHT for the single element is analytical, yielding a pole function $1/(\omega-z_j)$ with $z_j=\epsilon_j+i \delta_j$, with the sign convention for $\delta_j$ defined as discussed above according to time ordering.
Thus, choosing $b_j$ as in Eq.~\eqref{eq:1lor_basis} induces a SOP representation for $G$ according to Eq.~\eqref{eq:SOP}, with  
$A_i =a_i \in \mathbb{R}$. Once the SOP representation of $G$ is known, i.e. poles and amplitudes are known, the grid evaluation (inverse of the above) is trivial and amounts to performing the finite sum in Eq.~\eqref{eq:SOP}.
%
This approach ensures a full-frequency treatment of the propagator (approaching the continuum representation limit where Lorentzians becomes delta functions), while preserving 
an explicit knowledge of the analytical structure and continuation of $G$.

The main drawback of using Lorentzians to represent $G$ is related to the slowly decaying tails ($1/\omega^2$ for $\omega \to\infty$) induced in the spectral function when using finite broadening values $\delta_j$.
%
In order to improve on this, we introduce here $n$-th order generalized Lorentzians to obtain fast-decay basis functions. These are defined as
\begin{eqnarray}
    \abs{b_j(\omega)} = \mathcal{L}_{\delta_j}^n(\omega-\epsilon_j) = 
    \frac{1}{N_n\pi} \frac{\abs{\delta_j}^{2n-1}}{(\omega-\epsilon_j)^{2n}+(\delta_j)^{2n}},
    \label{eq:nlor_basis}
\end{eqnarray}
where 
$N_n = \left[n \sin(\frac{\pi}{2n})\right]^{-1}$ 
is the normalization factor (see Appendix~\ref{sec:SOP_nLor}).
The D-TOHT of $\mathcal{L}_\delta^n$ remains analytic and still yields a SOP representation for $G$ (see Appendix~\ref{sec:SOP_nLor}):
\begin{equation}
    \label{eq:TOHT_nlor_basis}
    \int_{\mathcal{C}} d\omega'\frac{\mathcal{L}_{\delta_j}^n(\omega'-\epsilon_j)}{\omega-\omega'-i0^+\,  \text{sgn}(\epsilon_j)} 
       =
    \sum^{n-1}_{m=0} \frac{ \alpha_{m}}{\omega-\zeta_{j,m}},
\end{equation}
with residues $\alpha_{m}$ and poles $\zeta_{j,m}$ given by
\begin{eqnarray}
\label{eq:TOHT_nlor_basis_res}
\alpha_{m} &=&\frac{1}{i N_n n} e^{i\frac{\pi}{2n}\left(1+2m\right)} \\[5pt]
\zeta_{j,m}&=&\epsilon_j+e^{i\frac{\pi}{2n}\left(1+2m\right)}\delta_j\, .
\label{eq:TOHT_nlor_basis_poles}
\end{eqnarray}

Importantly, $\alpha_{m}$ are complex (and so become the residues $A_i=a_j \alpha_{m}$  in the SOP representation, $i$ being a combined index).
Thus, the spectral function of this SOP has contribution by both the real and the imaginary part of each Lorentzian pole $1/(\omega-\zeta_{j,m})$, resulting in a overall faster decay than each single Lorentzian. Also, it is worth noting that, as for standard Lorentzians, a normalized $n$-th-order-Lorentzian approaches a Dirac delta for $\delta_j\to0^+$.
Owing to their fast decay and to this last property, using a SOP for $G_0$ in term of $n$-th Lorentzians provides a faster convergence for $\delta\to 0^+$, in comparison with a SOP representation built on ordinary Lorentzians, as will be also shown later. While the use of $n$-th order generalized Lorentzians to represent the spectral function $A(\omega)$ provides a faster decay in the imaginary-part of the propagator, it results in a multiplication of the number of poles in the SOP for $G$ (by the degree of the Lorentzian), and in having complex residues. As it will be shown in Sec.~\ref{sec:method_analytical}, the decay properties are fundamental for evaluating the moments of a SOP representation, assuring absolute convergence up to order $2(n-1)$. Also, the use of faster decay basis elements when representing the spectral function improves on the stability of the representation procedure, reducing the off-diagonal elements of the overlap matrix of the basis (see Sec.~\ref{sec:method_transform}).

Alternatively to $n$-th order Lorentzians, one could consider e.g. using Gaussian functions to represent $A(\omega)$, and consequently $G(\omega)$, as done in Refs.~\cite{von_barth_self-consistent_1996,holm_self-consistent_1997}.
Gaussians also allow for an analytical expression of the D-TOHT, at the price, though, of invoking the Dawson~\cite{Abramowitz1965book} or Faddeeva~\cite{virtanen_scipy_2020} functions to evaluate the real part of the propagator. Because of this, SOP expressions are not available, and basic operations involving propagators (such as those described in Sec.~\ref{sec:method_analytical}) cannot be evaluated analytically and need to be worked out in other ways, e.g. numerically or recasting the expressions in terms of propagator spectral functions~\cite{von_barth_self-consistent_1996}.

\subsection{Transform to a sum over poles}
\label{sec:method_transform}
Once the SOP representation has been introduced, the next important step is to determine numerically the SOP coefficients $A_i$ in Eq.~\eqref{eq:SOP}, given an evaluation of $G$ on a frequency grid.
According to the discussion of Sec.~\ref{sec:method_representation}, the SOP representation can be seen equivalently as a representation for the Green's function $G$ or for the spectral function $A$.

As a first case, we consider representing $A(\omega)$ according to Eq.~\eqref{eq:A_representation}, and we do it using the basis of $n$th-order generalized Lorentzians introduced in Eq.~\eqref{eq:nlor_basis}.
First, we obtain the coefficients $a_j$ of the representation by performing a non-negative-least-square (NNLS) fit~\cite{lawson_solving_1995,virtanen_scipy_2020}, thus assuring the positivity of all $a_j$. Then, we use Eqs.~(\ref{eq:TOHT_nlor_basis_res}-\ref{eq:TOHT_nlor_basis_poles}) to get the SOP representation for the propagator.
While the position and broadening $(\epsilon_j,\delta_j)$ of the $n$-th order Lorentzians could also be optimized by means of a non-linear NNLS fit, here we consider them centred at $\epsilon_j=\frac{1}{2}\left(\omega_j+\omega_{j-1}\right)$ and broadened with $\delta_j=\abs{\omega_j-\omega_{j-1}}$, and we just linearly optimize $a_j$. Also, for numerical reasons we prefer to work with the bare imaginary part of $G$, i.e. without imposing the sign factor of Eq.~\eqref{eq:G_to_A}, since this function is smoother than the actual spectral function $A(\omega)$ close to the Fermi level.

Alternatively, one could consider the basis representation induced on $G$ via Eq.~\eqref{eq:SOP} in order to directly obtain the $A_i$ and $z_i$ coefficients (residues and poles). As for $A(\omega)$, this can be achieved by a linear or non-linear LS fit (or interpolation) taking advantage of the knowledge of the whole $G(\omega)$ on a frequency grid (and not just of $A$).
Interestingly, the SOP representation in Eq.~\eqref{eq:SOP} is a special case of a Pad\`e approximant, written as the ratio of polynomials of order $N-1$ and $N$, respectively.
Because of this, one can exploit Pad\`e-specific approaches to determine ($A_i$,$z_i$), such as, for instance, Thiele's recursive scheme~\cite{Lee1996PRB}.
We found that this leads to a very efficient method when few tens of poles are considered, becoming numerically unstable beyond. Moreover, since the residues are not constrained to be real and positive ($A_i$ are actually complex), there is no control over the time-ordered position of the poles, and the procedure is non-trivial to extend to the case of $n$-th order Lorentzians. For the above reasons, in the present work we adopt the first approach, based on the representation of $A(\omega)$.

\subsection{Analytical expressions}
\label{sec:method_analytical}
Once the SOP representation of a dynamical propagator is available, a number of analytical expressions hold.
For instance, the convolution of propagators, such as those involved in the evaluation of the independent-particle polarizabilities in terms of the Green's functions, can be evaluated using Cauchy's residue theorem:
\begin{multline}
    \int_{-\infty}^{+\infty}  \frac{d\omega'}{2\pi i} \,G(\omega+\omega')\Tilde{G}(\omega')d\omega' = \\
    =
    \sum_{i,j} \int_{-\infty}^{+\infty}
    \frac{d\omega'}{2\pi i} \, \frac{A_i}{\omega+\omega'-z_i}\,\frac{\Tilde{A}_j}{\omega'-\Tilde{z}_j} \\
    =
    \sum_{\substack{i,j \\ \Im{z_i}<0 \\ \Im{\Tilde{z}_j}>0 }} \frac{A_i \Tilde{A}_j}{\omega+\Tilde{z}_j-z_i}
    - \sum_{\substack{i,j \\ \Im{z_i}>0 \\ \Im{\Tilde{z}_j}<0 }} \frac{A_i \Tilde{A}_j}{\omega +\Tilde{z}_j-z_i}.
\label{eq:convolutionSOP}
\end{multline}
%
Using the the SOP for $G$, the following integrals can also be computed explicitly: 
\begin{eqnarray}
    E_m[G]&=&
    \int_{-\infty}^{+\infty} \frac{d\omega}{2\pi i} 
    e^{i\omega0^+}
    \omega^m G(\omega)
    \nonumber \\[6pt]
    &=&\sum_{\substack{i \\ \Im{z_i}>0}} A_i z^{m}_i,
    \label{eq:moments}
\end{eqnarray}
where we refer to the term $E_m[G]$ as the $m$-th (regularized) moment of $G$. 
We restrict the discussion to the first $m=0$ and $m=1$ moments, since those are of interest for calculating the number of particles and the total energy in MBPT (see Sec.~\ref{sec:method_HEG_freqintegrated} for details). Higher order moments would require a stronger regularization factor in Eq.~\eqref{eq:moments} than $e^{i\omega0^+}$.
We underline that if one uses an $n$-th-order Lorentzian basis to represent $G(\omega)$ on SOP, the first $2(n-1)$ moments coincide with the moments of its occupied spectral function $\int_{-\infty}^{\mu} d\omega\, \omega^{2(n-1)} A(\omega)$. This is shown in Appendix~\ref{sec:Moments_and_occupiedM}.

\subsection{Numerical validation}
\label{sec:validation}
%

\begin{figure}
    \centering
    \includegraphics[width=\columnwidth]{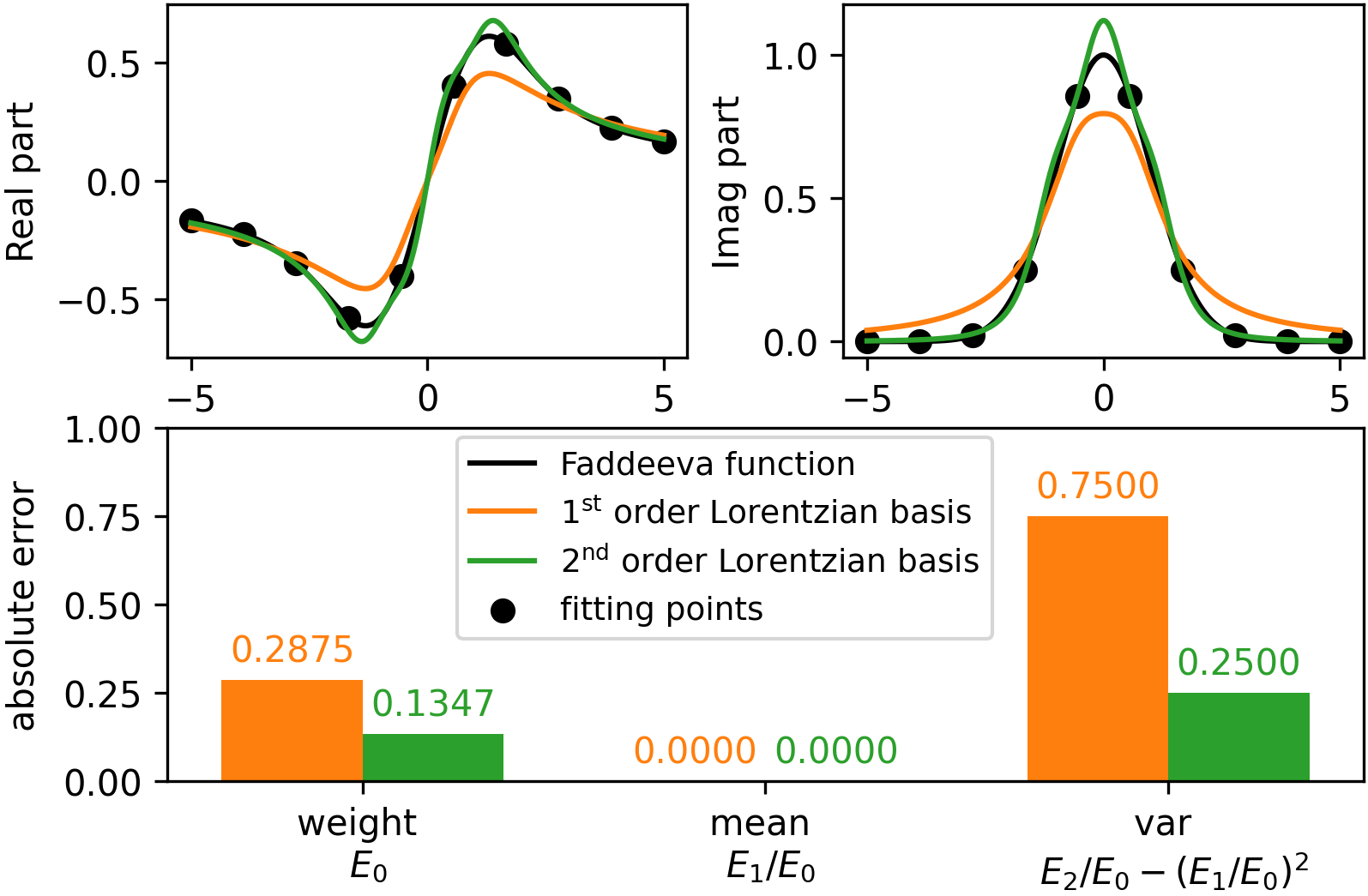}
    \caption{Numerical example of a transformation to a SOP form. Upper panels: the function to represent is chosen as the Faddeeva function (black line), sampled using only $10$ points (black dots). Following the strategy described in Sec.~\ref{sec:method_transform}, we represent the spectral function of the Faddeeva (a Gaussian) on $1^\text{st}$ order (orange) and $2^\text{nd}$ order (green) Lorentzians (centred on the midpoint between adjacent grid points, broadened with the size of the interval). The $10$ sampled points are fed to the NNLS fit to obtain the coefficients of both the $1^\text{st}$ and $2^\text{nd}$ Lorentzian basis, and Eqs.~\eqref{eq:TOHT_nlor_basis_res} and~\eqref{eq:TOHT_nlor_basis_poles} are used to get the SOP. Then, the resulting SOPs are plotted on a fine grid (continuous orange and green lines). Lower panel: absolute error on the computed 0-th to 2nd moments of the SOP representations (obtained in the upper panel) and the analytical Gaussian moments of the Faddeeva function.}
    \label{fig:fadeevaVsLorWithMeans}
\end{figure}

In the following we highlight numerically some properties of the SOP representation.
To this aim, we consider a propagator $G$ obtained as the Hilbert transform (HT) of a Gaussian, analytically expressed via the Faddeeva~\cite{virtanen_scipy_2020} function (black curves in the top panels of Fig.~\ref{fig:fadeevaVsLorWithMeans}). Here we have assumed the Fermi level to be far enough from the imaginary part of $G$ such that the retarded HT can be used. 
The objective of the validation is to transform the Faddeeva Green's function sampled on a finite grid to a SOP representation. Following Sec.~\ref{sec:method_transform}, we represent the Gaussian spectral function on first (ordinary) and second-order Lorentzians, use Eqs.~\eqref{eq:TOHT_nlor_basis_res} and~\eqref{eq:TOHT_nlor_basis_poles} to get the resulting SOP representation, and then compare the results ---orange and green lines for $1^\text{st}$ and $2^\text{nd}$ order basis, respectively--- with the starting Faddeeva Green's function calculated on a much finer grid. We choose to feed the NNLS fitting algorithm with $10$ sampling points (for the imaginary part of $G$) and to use $9$ basis functions centered at the midpoints of the grid, and broadened with the width of the interval (in order to ensure an exhaustive cover of the domain). Then we use  Eqs.~(\ref{eq:TOHT_nlor_basis}-\ref{eq:TOHT_nlor_basis_poles}) to obtain the SOP representation of $G$ (in the case of $n$th order Lorentzians) from the output of the NNLS procedure.

Underscoring the quality of the representation, the upper panels of Fig.~\ref{fig:fadeevaVsLorWithMeans} show that using faster-decay $2^\text{nd}$ order Lorentzians provides a more accurate result for both the real (left) and imaginary (right) part of $G$. In the lower panel we also compare the moments of the Faddeeva function with those obtained from Eq.~\eqref{eq:moments} and Eq.~\eqref{eq:occupiedMoments}. Since absolute convergence for the first and second moments is not ensured for $1^\text{st}$ order Lorentzians, meaning that Eq.~\eqref{eq:occupiedMoments} does not hold, $E_{n>1}[G]$ has to be calculated according to Eq.~\eqref{eq:moments} and is in general complex. In order to obtain a meaningful result we take its real part, and consider the imaginary one as an error that must be controlled 
by extending the basis set towards completeness. 

\subsection{Algorithmic inversion on SOP}
\label{sec:Dyson_SOP}
As anticipated in the introduction, within the SOP approach the exact  solution at all frequencies of the Dyson equation can be remapped into the diagonalization of a static effective Hamiltonian (Hermitian only under special conditions), a procedure that we refer to as ``algorithmic-inversion method on sum over poles'' (AIM-SOP); this is a central result for the present work. 
As mentioned we will use the HEG as a paradigmatic test case, leaving the treatment of the non-homogeneous case to later work. Suppressing then the $k$ momentum index for simplicity, let us suppose to have the SOP representation of the self-energy $\Sigma(\omega)$ and the non-interacting Green's function $G_0(\omega)$ given by 
\begin{equation}
\Sigma(\omega)=\sum_{i=1}^{N}\frac{\Gamma_i}{\omega-\sigma_i}, \qquad G_0=\frac{1}{\omega-\epsilon_0}.
\end{equation}
Taking advantage of these expressions, the Dyson equation can be rewritten as
\begin{eqnarray}
     G(\omega)&=&\left[G_0^{-1}(\omega)-\Sigma(\omega)\right]^{-1}
     =\frac{1}{\omega-\epsilon_0-\Sigma(\omega)}
     \nonumber \\
     &=&
     \frac{(\omega-\sigma_1)\cdots(\omega-\sigma_N)}{T_N(\omega)},
     \label{eq:Dyson_G}
\end{eqnarray}
in which the $N+1$ roots of the polynomial
\begin{eqnarray}
     T_N(\omega)&=&
     (\omega-\epsilon_0)
     \!\!\! \prod_{i=1,N} \!\! (\omega-\sigma_i) 
     \nonumber \\
     &-& \sum_{j=1,N} \Gamma_j 
     \prod_{\substack{i=1,N\\i\neq j}} (\omega-\sigma_i)
\end{eqnarray}
are the $N+1$ poles of the Green's function (as expected when the self-energy has $N$ poles). Then, the key statement of this Section is that the roots of $T_N$ can be obtained as the eigenvalues of the $\left(N+1\right)\times \left(N+1\right)$ matrix
\begin{equation}
    H_\mathrm{AIM}=\begin{pmatrix}
        \epsilon_0 & \sqrt{\Gamma_1} & \dots & \sqrt{\Gamma_N}\\
        \sqrt{\Gamma_1} & \sigma_1 & 0 & 0 \\
        \vdots & 0 & \ddots & 0\\
        \sqrt{\Gamma_N} & 0 & \dots & \sigma_N
    \end{pmatrix}.
    \label{eq:AIM_matrix}
\end{equation}
We prove this statement by observing that the characteristic polynomial of $H_\mathrm{AIM}$ is $T_N(\omega)$, and we proceed by induction. Since the $N=1$ case is trivial we move to the $N$-th case: using the Laplace expansion on the last line, the characteristic polynomial of the $N$-th case can be written as 
%
%
\begin{eqnarray}
    p_{H_\mathrm{AIM}}(\omega)&=& 
      \begin{vmatrix}
       \omega-\epsilon_0 & -\sqrt{\Gamma_1} & \dots & -\sqrt{\Gamma_N}\\
        -\sqrt{\Gamma_1} & \omega-\sigma_1 & 0 & 0 \\
        \vdots & 0 & \ddots & 0\\
        -\sqrt{\Gamma_N} & 0 & \dots & \omega-\sigma_N
     \end{vmatrix}
  \nonumber \\[5pt]
  &=&(\omega-\sigma_N) T_{N-1}(\omega) +(-1)^{N} \sqrt{\Gamma_N}\times 
  \nonumber \\[5pt]
  &\times&
    \begin{vmatrix}
        -\sqrt{\Gamma_1} & \dots & -\sqrt{\Gamma_{N-1}} &  -\sqrt{\Gamma_N}\\
        \omega-\sigma_1 & 0 & 0& 0 \\
        0 & \ddots & 0 & 0\\
        0 & \dots & \omega-\sigma_{N-1} & 0
    \end{vmatrix}
\end{eqnarray}
where we have used the induction hypotheses in the first term of the rhs. Applying the same procedure to the last column of the second term we obtain
\begin{eqnarray}
    p_{H_\mathrm{AIM}}(\omega)&=&(\omega-\sigma_N) T_{N-1}(\omega)  
    -\Gamma_N \!\!\!\prod_{i=1,N-1}(\omega-\sigma_i)
    \nonumber \\
    &=&T_N(\omega),
\end{eqnarray}
which completes the proof.

Calling $\epsilon_i$ the poles of $G$ we calculate the residues by equating
\begin{equation}
    G(\omega)=\sum_{i=1}^{N+1}\frac{A_i}{\omega-z_i}=
    \frac{(\omega-\sigma_1)\cdots(\omega-\sigma_N)}
    {(\omega-z_1)\cdots(\omega-z_{N+1})},
\end{equation}
and performing the limit $\lim_{\omega\to z_i}(\omega-z_i)$ on both sides (Heaviside cover-up method~\cite{thomas_calculus_1988}), obtaining:
\begin{equation}
  A_i = \frac{\prod_{k=1}^N (z_i-\sigma_k)}
    {\prod_{j=1,\,j\neq i}^{N+1} (z_i-z_j)}.
    \label{eq:amps_AI}
\end{equation}
We have thus proven that by knowing $\Sigma$ represented on SOP, the SOP expression of $G$ can be found by the diagonalization of the AIM-SOP matrix $H_\mathrm{AIM}$ followed by the evaluation of the residues using Eq.~\eqref{eq:amps_AI}.

It is worth noting that the $H_\mathrm{AIM}$ matrix becomes (or can be made) Hermitian under special conditions. This happens when the self-energy residues $\Gamma_i$ are real and positive, and the self-energy poles have all the same imaginary part $i\delta$, with the usual time-ordered convention according to Eq.~\eqref{eq:SOP}, also equal to the broadening assumed for the $G_0$ pole. Then it is possible to include the imaginary part of the poles in the frequency variable $\omega$, and invert $G(\Tilde{\omega} \in \gamma)$ on this time-ordered-complex path, in order to have $H_\mathrm{AIM}$ with only the real part of the poles along the diagonal. Finally, in order to have $G(\omega \in \mathcal{R})$ we analytically continue the solution to the real axis, obtaining $\Im{\epsilon_i}=\delta_i$ for the SOP of the Green's function.

We also stress that, given a self-energy represented on SOP, the solution provided by the algorithmic-inversion procedure is exact at all frequencies. This ensures the Green's function fulfills all the sum-rules implied by the Dyson equation, including e.g. the normalization of the spectral weight, and the first and second moments sum rules of the spectral function derived in Ref.~\cite{von_barth_self-consistent_1996}.
This result is crucial when evaluating frequency-integrated quantities of a Green's function, such as the number of particles or the total energy (see Sec.~\ref{sec:method_HEG_freqintegrated}).

Besides the solution of the Dyson equation for $G$, the AIM-SOP 
can also be used to solve the Dyson equation for the screened Coulomb interaction $W(\omega)$, 
\begin{eqnarray}
    \nonumber
    W(\omega) &=& v_c + v_c P(\omega) W(\omega) \\
              &=& \frac{1}{1-v_c P(\omega)}v_c = \epsilon^{-1}(\omega)v_c,
              \label{eq:Dyson_w}
\end{eqnarray}
i.e. to compute the SOP representation of $W(\omega)$ once a SOP for the irreducible polarizability $P(\omega)$ is provided. Here $v_c$ is the Coulomb potential (recalling that we are suppressing the momentum dependence for simplicity).
%
%
%
By letting $P(\omega)=\sum_i \frac{S_i}{\omega-g_i}$, we can write:
\begin{eqnarray}
  \nonumber
  \omega v_c P(\omega)&=&\sum_i \omega\frac{v_c{S}_i}{\omega-g_i}
  =v_c \sum_i S_i
  +\sum_i \frac{v_c g_i{S}_i}{\omega-g_i} \\
  &:=&c_0-C(\omega),
\end{eqnarray}
and following Eq.~\eqref{eq:Dyson_w} (multiplied by $\frac{\omega}{\omega}$) we have
\begin{equation}
    \epsilon^{-1}(\omega)=\frac{1}{1-v_c P(\omega)}=\frac{\omega} {\omega-c_0-C(\omega)}, 
    \label{eq:poles_epsm1}
\end{equation}
for which the AIM-SOP matrix can be used to find the poles of $\epsilon^{-1}(\omega)$ and $W(\omega)$. 
The amplitudes of $W$ are easily found using Eq.~\eqref{eq:amps_AI}. Note that by multiplying $\epsilon^{-1}$ by $\frac{\omega}{\omega}$ in Eq.~\eqref{eq:poles_epsm1} we have inserted an extra pole (at $\omega=0$) which we need to discard from the eigenvalues of the AIM matrix [before applying the residue formula of Eq.~\eqref{eq:amps_AI}], since it simplifies with the $\omega$ at the numerator of Eq.~\eqref{eq:poles_epsm1}.
Consequently $P$, $\epsilon^{-1}$, and $W$ all have the same number of poles, at variance with the solution of the Dyson equation for $G$, where the number of poles of $G$ is increased by one with respect to those of $\Sigma$.

\begin{figure}
    \centering
    \includegraphics[width=\columnwidth]{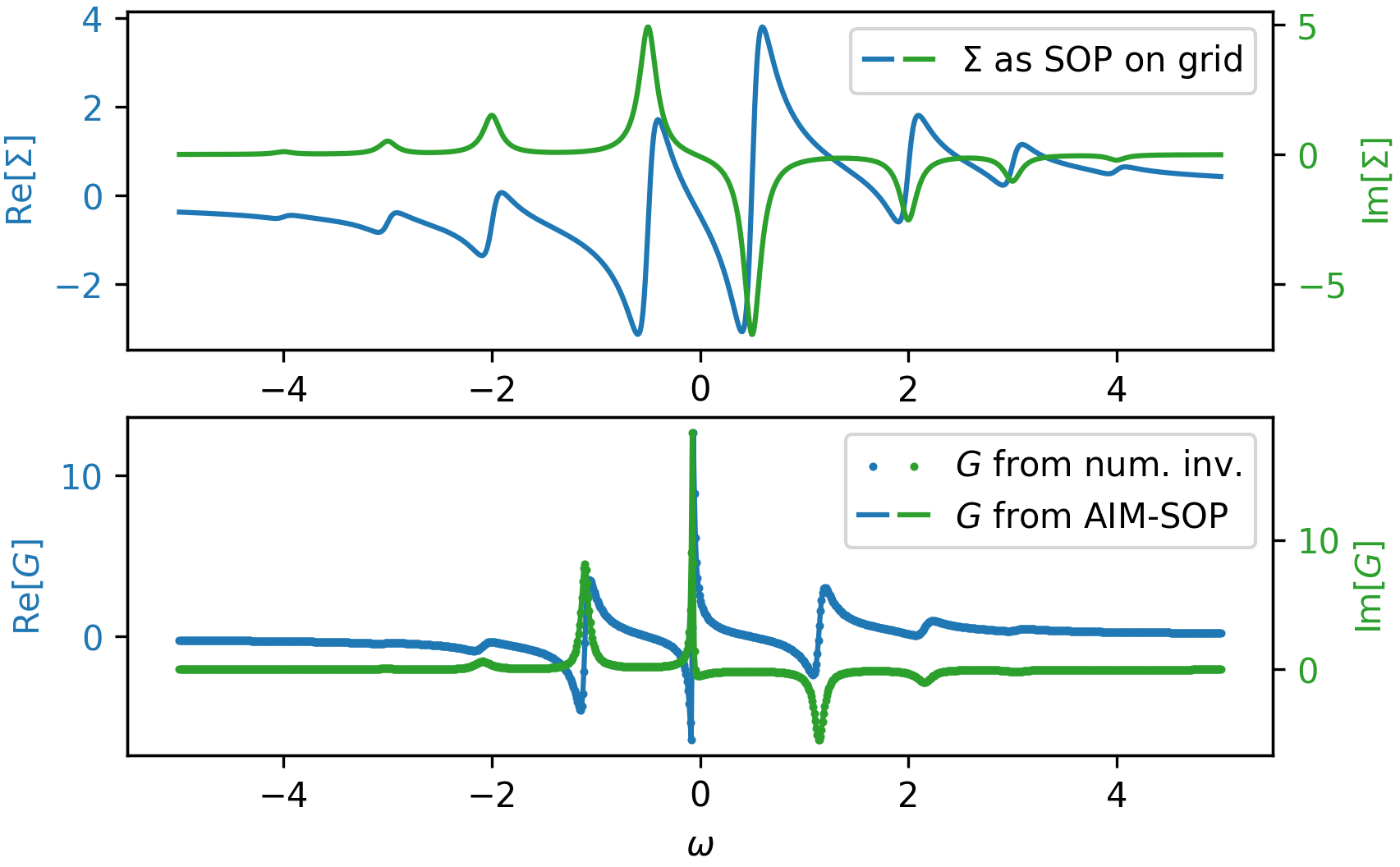}

    \caption{Numerical example for the algorithmic inversion method on sum over poles. Upper panel: Real (blue) and imaginary (green) part of a time-ordered self-energy, including 8 poles (the occupied pole of $G_0$ is not shown). Lower panel: Dyson-inverted propagator $G$ obtained with a numerical inversion on a grid (dotted) compared with the SOP representation obtained using AIM-SOP and evaluated on the same grid (solid line). Same color code for real and imaginary parts as in the upper panel.}
    \label{fig:AI_example}
\end{figure}

As a numerical test for AIM-SOP, we consider the Dyson equation for $G$ within the example of a time-ordered self-energy built with $8$ poles, as shown in the upper panel of Fig.~\ref{fig:AI_example} (the single pole of $G_0$ is not reported).
In the lower panel we compare the Green's function $G$ obtained from the numerical Dyson inversion on grid --- done evaluating $\Sigma$ on grid, and then inverting --- against the Green's function with the algorithmic inversion and evaluated on the frequency grid.
The results are identical at the precision of the calculated eigenvalues of the AIM-SOP matrix, since the amplitude calculation of Eq.~\eqref{eq:amps_AI} is typically very well conditioned. Notably, this procedure has been tested in cases where hundreds of poles are used for the self-energy, without any numerical instabilities. 


\section{Application: One-shot $G_0W_0$ in the HEG from AIM-SOP}
\label{sec:method_HEG}
%
For validation, we apply the AIM-SOP approach to the paradigmatic case of the homogeneous electron gas (HEG), treated at the $G_0W_0$ level of theory~\cite{aryasetiawan_thegwmethod_1998,reining_gw_2018,Martin-Reining-Ceperley2016book}. Since we calculate propagators on the real axis we can easily access spectral (frequency-dependent) properties. The calculation of frequency-integrated ground-state quantities (occupation numbers, total energies, and thermodynamic quantities in general)
can be obtained directly from the SOP representation of the spectral quantities computed in the procedure. We stress that usually~\cite{rojas_space-time_1995,rieger_gw_1999,garcia-gonzalez_self-consistent_2001} thermodynamic properties are obtained via additional calculations of propagators (e.g. on the imaginary axis), while in this work spectral properties and integrated quantities are obtained simultaneously using the SOP representation of propagators computed on the real axis.

While some quantities computed using the free-propagator $G_0$ have known analytical expressions, as is the case for the irreducible polarizability $P_0$ expressed via the Lindhard function~\cite{Fetter-Walecka1971book,giuliani_quantum_2005},
here we recompute explicitly all the propagators needed to evaluate the GW self-energy, making the treatment suitable also for self-consistent calculations. Therefore in the following the only assumption we make is to consider the Green's function as represented on SOP.

\subsection{HEG propagators on the real frequency axis}
\label{sec:method_HEG_propagators}

\begin{figure*}
    \centering
    \includegraphics[width=0.9\textwidth]{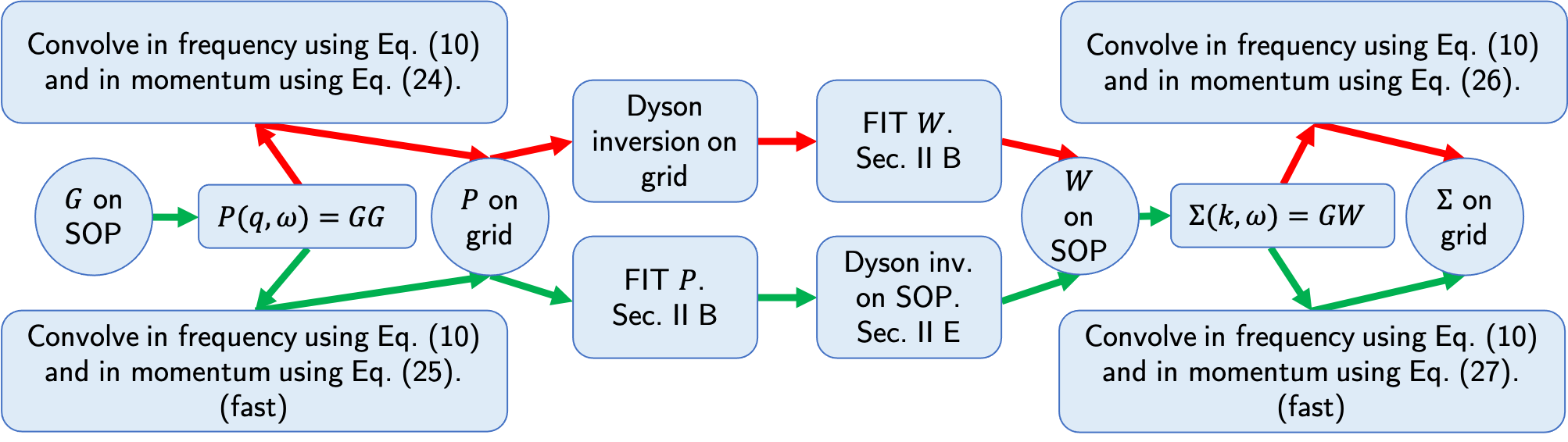}
    \caption{Flow chart representing different strategies for the calculation of the self-energy given a Green's function $G$ on SOP as input for the \texttt{heg\_sgm.x} code. The strategy used in this article is highlighted with green lines.
    }
    \label{fig:flow_charts_spectral}
\end{figure*}

In order to solve a one-shot G$_0$W$_0$  cycle for the spin-unpolarized HEG, we first need to compute the irreducible polarizability at the independent-particle (or RPA) level, according to the integral
\begin{multline}
    P(q,\omega)=2\int \frac{d\mathbf{k}}{(2\pi)^3} 
    \int 
    \frac{d\omega'}{2\pi i} \, G(\abs{\mathbf{k+q}},\omega+\omega')G(k, \omega'),
    \label{eq:polarizabilityIntStandard}
\end{multline}
where $k=|\mathbf{k}|$ and $q=|\mathbf{q}|$ are the moduli of the electron and transferred quasi-momenta, respectively.  
To compute Eq.~\eqref{eq:polarizabilityIntStandard}, the frequency integral (convolution) is performed analytically according to Eq.~\eqref{eq:convolutionSOP}. Then we integrate numerically in spherical coordinates by performing the variable change $x=\abs{\mathbf{k+q}}$ on the azimuthal angle of $\mathbf{k}$,
\begin{multline}
    P(q,\omega)=\frac{2}{q(2\pi)^2}  \int_0^{+\infty} dk \, k \int_{\abs{k-q}}^{\abs{k+q}} dx \, x \\
    \times \int
    \frac{d\omega'}{2\pi i} \, G(x,\omega+\omega')G(k, \omega'),
    \label{eq:polarizabilityInt}
\end{multline}
which allows for the pre-calculation of the analytical convolutions on the two-dimensional $(x,k)$ grid, instead of on the three-dimensional $(k,q,\theta)$ space. Exploiting the parity of $P(\omega)$, it is also possible to limit the $\mathbf{k}$ integration to the occupied states (see Appendix~\ref{sec:parity_polariz_integral}). The numerical integration on the momentum is performed using the trapezoidal rule, which ensures exponential convergence for decaying functions~\cite{trefethen_exponentially_2014}.

In order to have a SOP representation for the screened potential $W$ we transform the polarizability calculated on a frequency grid (at fixed momentum $q$) to a SOP performing a NNLS fitting, following the procedure of Sec.~\ref{sec:method_transform}. We then solve the Dyson equation using the algorithmic inversion for the polarizability (see Sec.~\ref{sec:Dyson_SOP}) to obtain a SOP for $W$, and use it for the $GW$ integral. 
An alternative possibility would be to solve the Dyson equation on a grid (which, due to homogeneity, is an algebraic inversion), and then transform $W$ to a SOP representation.
Even admitting for an exact interpolation for the SOP of $W$ on the calculated frequencies (where the Dyson equation is solved on grid), this SOP would suffer from not having solved the Dyson equation for all other frequencies. 
Very differently, the SOP obtained from the algorithmic inversion provides for an exact solution of the Dyson equation at all frequencies (see Sec.~\ref{sec:Dyson_SOP}). 
Thus, the sum rules implied by the Dyson equation (moments of the spectral function) are all obeyed by the SOP obtained from the algorithmic inversion, being the exact solution at all frequencies. Conversely, this is not true for the grid inversion where the solution is exact only for isolated frequencies.

Concerning the self-energy integral
\begin{multline}
    \Sigma(k,\omega)=\Sigma_x(k)+\frac{1}{(2\pi)^3}\int d\mathbf{q} \\
    \times \int_{-\infty}^{+\infty} \frac{d\omega'}{2\pi i} \, G(\abs{\mathbf{k+q}},\omega+\omega')W_\mathrm{corr}(q,\omega'),
    \label{eq:sigmaIntegralStandard}
\end{multline}
where $W_\mathrm{corr}=W-v_c$, we can still use Eq.~\eqref{eq:convolutionSOP} since we have the SOP representation of $W$. Again, in Eq.~\eqref{eq:sigmaIntegralStandard} we perform the $x=\abs{\mathbf{k+q}}$ change of variable obtaining
\begin{multline}
    \Sigma(k,\omega)=\Sigma_x(k)+\frac{1}{k(2\pi)^2}  \int_0^{+\infty} dq \, q \int_{\abs{k-q}}^{\abs{k+q}} dx \\
    \times \int
    \frac{d\omega'}{2\pi i} \, G(x,\omega+\omega')W_\mathrm{corr}(q,\omega'),
    \label{eq:sigmaIntegral}
\end{multline}
which allows for fewer convolutions (as for the polarizability integral), and use trapezoidal weights as in Eq.~\eqref{eq:polarizabilityInt} for the momentum integration. The solution of the Dyson equation for the Green's function using the algorithmic inversion, and the calculation of frequency-integrated (thermodynamic) quantities, are discussed in the next section.

In Fig.~\ref{fig:flow_charts_spectral} we show the overall flow chart describing the process of going from the knowledge of the initial Green's function to the calculation of the corresponding self-energy (for the HEG in the GW approximation), as implemented in the~\texttt{heg\_sgm.x} program of the \texttt{AGWX} suite~\cite{agwx-code}, by means of the SOP approach. As opposed to the path in red, where the Dyson equations are solved on grids, in the green path we highlight the protocol followed in the present work. The crucial difference between the two approaches is the use of the algorithmic-inversion method in order to solve exactly the Dyson equation, providing a SOP for $W$ obeying all sum rules implied by the Dyson equation, as previously discussed in this Section.


\subsection{Frequency-integrated quantities and thermodynamics} \label{sec:method_HEG_freqintegrated}

\begin{figure*}
    \centering
    \includegraphics[width=0.9\textwidth]{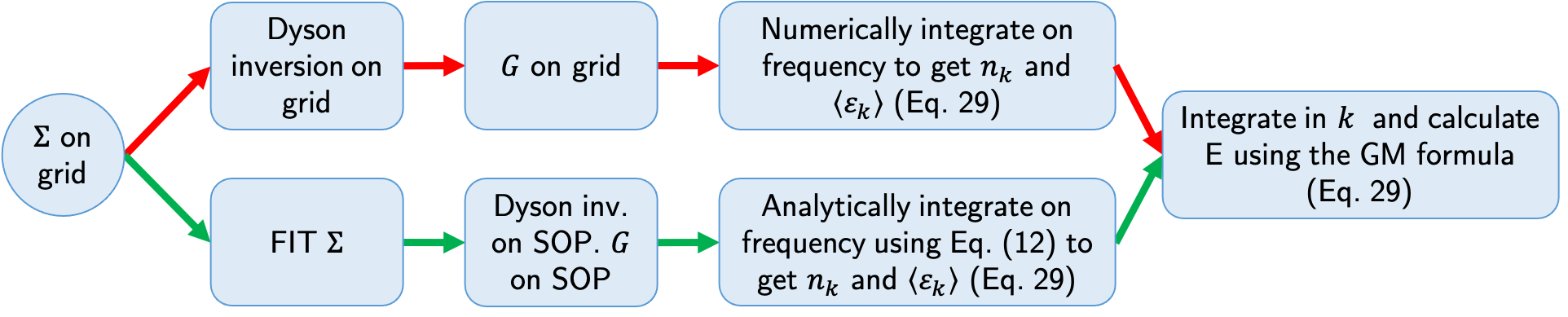}
    \caption{Flow chart representing different strategies for the calculation of the total-energy given a self-energy $\Sigma$, or a spectral function $A$, on a frequency grid as input for the \texttt{heg\_sgm.x} code. The strategy used in this article is highlighted with green lines.}
    \label{fig:flow_charts_total_ene}
\end{figure*}

Having obtained the self-energy on a frequency grid following the procedure described in Sec.~\ref{sec:method_HEG_propagators}, we evaluate the Green's function together with some related frequency-integrated quantities. 
As mentioned, the SOP approach plays here a central role, enabling the possibility of performing analytical integrals for the moments of $G$, as those involved in the Galitskii-Migdal expression for the total energy [see Eq.~\eqref{eq:Galitski-Migdal} below], and thus to have accurate thermodynamic (frequency-integrated) quantities. Moreover, the use of the algorithmic inversion allows for the exact solution the Dyson equation for the Green's function at all frequencies. The conservation of all sum rules (implied by the Dyson equation, see Sec.~\ref{sec:method_SOP}) guaranteed by the AIM-SOP is fundamental when calculating the occupied moments of the spectral function. As an example, the normalization condition of the spectral function is automatically satisfied when $G$ on SOP is obtained using the algorithmic inversion, and allows for not having fitting constrains which would be required, e.g., if we were to use a grid inversion.

In order to exploit the AIM-SOP to get $G$ on a SOP, we obtain the SOP representation of the self-energy by performing a NNLS fitting of $\text{Im}{\Sigma}(\omega)$ (see Sec.\ref{sec:method_transform}). 
%
Then, in order to compute the total energy from the knowledge of the Green's function $G$, we use the 
Galitski-Migdal expression~\cite{Fetter-Walecka1971book,Martin-Reining-Ceperley2016book},
\begin{eqnarray}
    \frac{E}{V}&=&\int \frac{d\mathbf{k}}{(2\pi)^3} \left[ \int_{-\infty}^{\mu} d\omega\ \omega A(k,\omega) +
    \frac{k^2}{2}\int_{-\infty}^{\mu}d\omega A(k,\omega) \right]
    \nonumber \\
    &=&\int \frac{d\mathbf{k}}{(2\pi)^3} \bigg[ \expval{\epsilon_k} +  \frac{k^2}{2} n_k \bigg],
\label{eq:Galitski-Migdal}
\end{eqnarray}
here in Hartree units, where $V$ is the volume of the periodic cell of the electron gas.
In this expression, the frequency integrals are performed using the SOP for $G$, and exploiting  Eq.~\eqref{eq:moments} with $m=1$ and $m=0$ for the first and second terms, respectively. 
Here $n_k$ is the $k$-resolved occupation function, which sums to the total number of particles when integrated over momentum, and $\expval{\epsilon_k}$ is the occupied band, i.e. the first momentum of the occupied spectral function.
For both $m=0$ and $m=1$ moments, the equality between the moments of the Green's function and the moments of the occupied spectral function, Eq.~\eqref{eq:moments} and Eq.~\eqref{eq:occupiedMoments}, is assured by having used the algorithmic inversion when obtaining the SOP for the Green's function. Indeed, the knowledge of the self-energy on SOP and the use of the algorithmic inversion for solving exactly the Dyson equation ensures that the spectral function
\begin{equation}
A=\frac{1}{\pi}\frac{\abs{\text{Im}{\Sigma(\omega)}}}{[\omega-\epsilon_0-\text{Re}{\Sigma(\omega)}]^2+[\text{Im}{\Sigma(\omega)}]^2},
\end{equation}
decays at least as $\frac{\text{Im}{\Sigma}}{\omega^2}=o(\omega^{-3})$, thereby making the first two occupied moments (see Sec.~\ref{sec:method_analytical}) converge.

Similarly to the discussion in Sec.~\ref{sec:method_HEG_propagators}, the SOP approach combined with the algorithmic inversion allows one to follow the workflow highlighted by the green path in Fig.~\ref{fig:flow_charts_total_ene}. Overall, the results presented Sec.~\ref{sec:results} are obtained using an implementation of the above approach in the~\texttt{heg\_sgm.x} program of the \texttt{AGWX} suite~\cite{agwx-code}.

\subsection{Numerical details} 
\label{sec:numerical_details}

\begin{figure}
    \centering
    \includegraphics[width=\columnwidth]{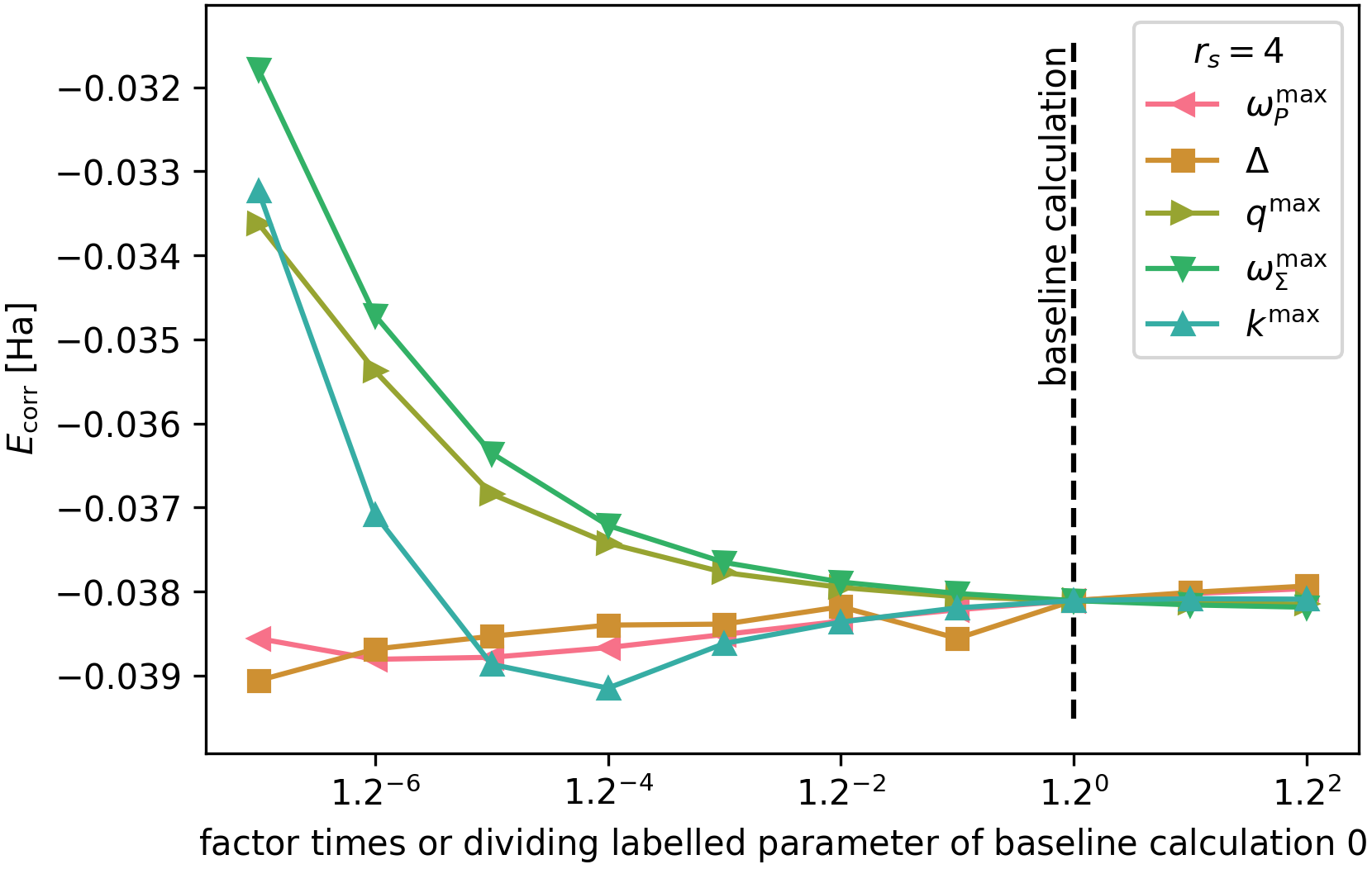}
    \caption{Convergence study for the correlation energy per particle $E_\mathrm{corr}$, obtained with the Galitzki-Migdal formula, and using a Green function from a $G_0W_0$ calculation for the HEG at $r_s=4$. The parameters to converge are explained in Sec.~\ref{sec:numerical_details}. We choose to converge $E_\mathrm{corr}$ for each parameter taking all the others fixed at the converged (second to last point) value. For each different parameter, we increase step-by-step its value by $20\%$ in the convergent direction.}
    \label{fig:totEnecG0W0rs4}
\end{figure}

Here we discuss and report the parameters that control the numerical accuracy of the quantities (polarizability, self-energy, total energy) computed by means of Eqs.~\eqref{eq:polarizabilityInt},~\eqref{eq:sigmaIntegral} and~\eqref{eq:Galitski-Migdal}. In practice, this corresponds to going from left to right in the flow diagram of Fig.~\ref{fig:flow_charts_spectral} following the green path, performing all calculations mentioned in the boxes.
The first quantity to be computed is the polarizability $P(q,\omega)$. For each momentum $q$ and frequency $\omega$, we perform the integral of Eq.~\eqref{eq:polarizabilityInt}. As $k$ in the integral is limited by $k_f$ (see Sec.~\ref{sec:method_HEG_propagators}), 
the discretization of the $k$- and $x$-grids, $\Delta k_P$ and $\Delta x_P$, has to be converged to the zero spacing limit. 
Also, it is necessary to converge to zero the spacing of the momentum and frequency points of the polarizability-$(q,\omega)$ grid, controlled by $\Delta q$ and $\Delta {\omega_P}$, along with the grid-upper limits (to infinity) $q^\mathrm{max}$ and $\omega^\mathrm{max}_{P}$. 

Moving to the central part of the flow chart in Fig.~\ref{fig:flow_charts_spectral}, the SOP representation of the polarizability is obtained following the method of Sec.~\ref{sec:method_transform}, and placing the center of the $2^\text{nd}$ order Lorentzians on the mid points of the frequency grid, which improves the accuracy of the fit as $\Delta {\omega_P}\to0$. Next, we employ the algorithmic-inversion method to go from the SOP representation of the polarizability to the SOP of the screened-potential $W$ 
(exact to machine precision, see Sec.~\ref{sec:Dyson_SOP}). Using the SOP representation of $W$ (and of $G$), the self-energy integral (right part of  Fig.~\ref{fig:flow_charts_spectral}), Eq.~\eqref{eq:sigmaIntegral},
is formally identical to the integral in Eq.~\eqref{eq:polarizabilityInt} for the polarizability. Therefore, the remaining parameters to converge are $\Delta x_{\Sigma}$, $\Delta k$, $\Delta {\omega_\Sigma}$, $k^\mathrm{max}$, and $\omega^\mathrm{max}_{\Sigma}$ (using the same notation adopted above). As for the screened-potential $W$, we obtain the SOP representation of the self-energy following Sec.~\ref{sec:method_transform}, and placing $2^\text{nd}$-order Lorentzians on the mid points of the frequency grid. Finally, we obtain the SOP representation of the Green's function employing the algorithmic-inversion method.

In principle, for each computed quantity which depends on the Green's function $G$, e.g. via the spectral function or its integrals, we should study the numerical stability of the computational procedure with respect to all the above parameters. Our numerical approach allows for the evaluation of the Green's function and the related spectral quantities on the real-axis, which are then used for the computation of thermodynamic quantities.
In this work, we choose to converge the total energy (as obtained in Sec.~\ref{sec:numerical_details}), which is sensitive enough to guarantee a reasonable convergence for the other (spectral) properties of interest here. 
By changing individually each parameter (increase or decrease by $20\%$ of its value towards convergence), we study the stability of the total energy against the selected parameter, keeping the values of all the others fixed at a reference point (baseline calculation of Fig.~\ref{fig:totEnecG0W0rs4}). Each target parameter is then converged separately 
until a plateau for the subsequent values of the computed quantity is observed. We evaluate the error on the result considering the two most distant values among those in the plateau.

Importantly, it is possible to reduce the number of parameters to converge from $13$ to $5$, by linking all the grid-spacing and broadening parameters together into a single variable, $\Delta$, which ensures convergence for $\Delta\to0^+$. Specifically, we bind those parameters together by setting $\Delta = \Delta k_P = 5\Delta x_{P} = \frac{1}{6}\Delta\omega_{W} = \frac{1}{9}\Delta q  =\frac{1}{25} \Delta\omega_{\Sigma} = \Delta x_{\Sigma} = \frac{1}{3}\Delta k_{\Sigma}=\frac{5}{4}\delta_P=\frac{1}{100}\delta_\Sigma$. 
Together with $\Delta$, the grid-limit parameters are converged separately, following the strategy designed above. The converged values obtained for all the calculated densities are: $\Delta = 0.004\ k_f$, $q^\mathrm{max} = 7.292\ k_f$, $k^\mathrm{max} = 3.60\ k_f$, $\omega^\mathrm{max}_{P} = 5.0\ \epsilon_f$, $\omega^\mathrm{max}_{\Sigma} = 10.985\ \epsilon_f$, where $k_f$ is the Fermi momentum and $\epsilon_f$ the Fermi energy.

\section{Results}
\label{sec:results}
%
In this Section we discuss the results obtained applying the SOP approach to the case of the one-shot $G_0W_0$ calculation in the HEG. First we extensively discuss the $r_s=4$ case, also one of the most studied in the literature, then in Sec.~\ref{sec:results_HEG_GoWo} we provide the results for densities ranging from $r_s=1$ to $r_s=10$.

\subsection{Spectral propagators on the real axis}
\label{sec:results_HEG_propagators}
%

\begin{figure}
    \centering
    \includegraphics[width=\columnwidth]{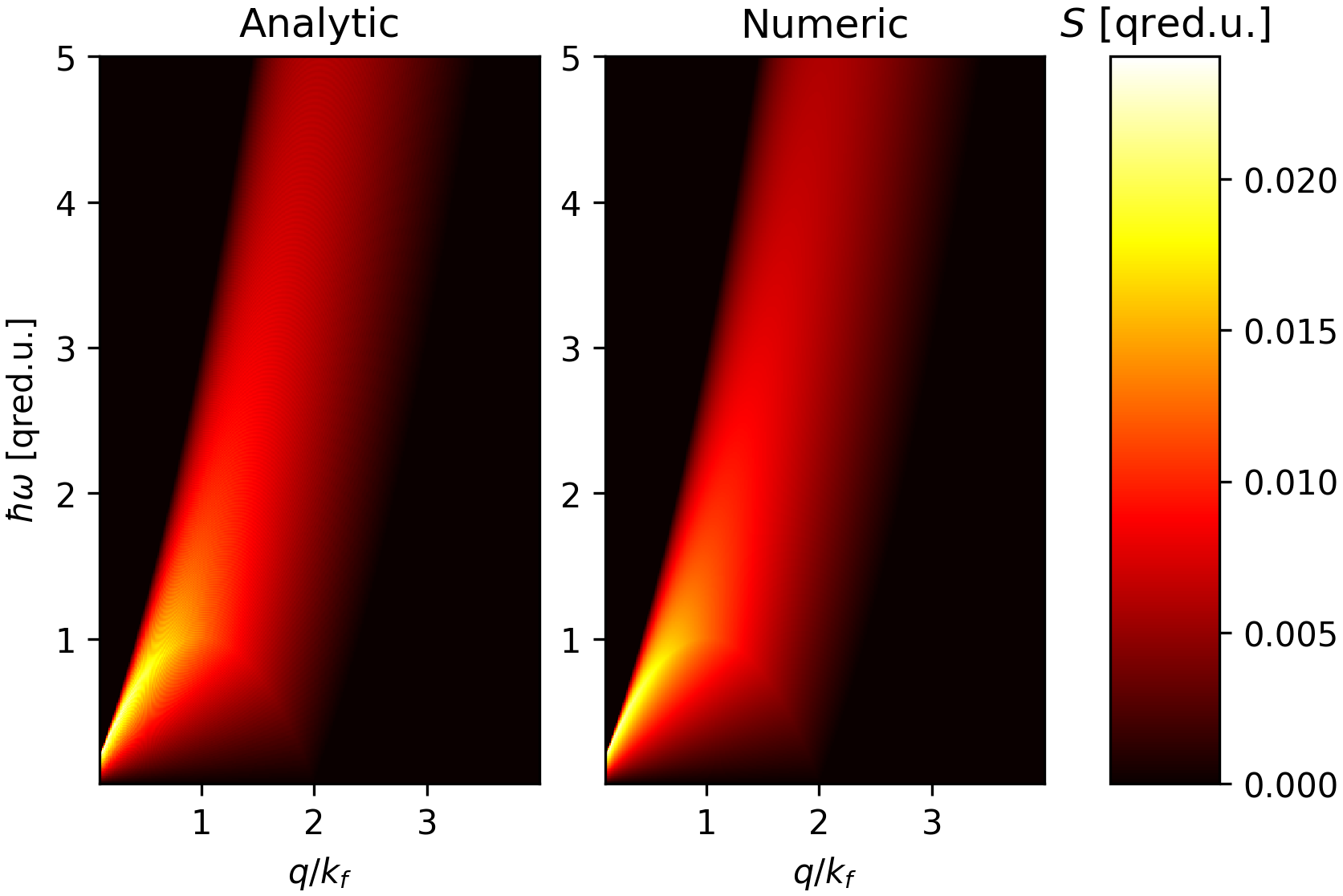}
    \caption{Spectral part of the polarizability of the HEG at $r_s=4$ in a one-shot $G_0W_0$ calculation. Energy units are $q$-reduced: $E/(\epsilon_f f(q/k_f))$ with $f(x)=x(1/\sqrt{x}+x)$ and $k_f$ the Fermi momentum. Left panel: data calculated with Eq.~\eqref{eq:polarizabilityInt}, represented in SOP, and then evaluated on a frequency grid. Right panel: Analytic results~\cite{fetter_quantum_2003} on the same frequency-momentum grid.}
    \label{fig:polarizabilityPlotCalsVsTheory}
\end{figure}

\begin{figure}
    \centering
    \includegraphics[width=\columnwidth]{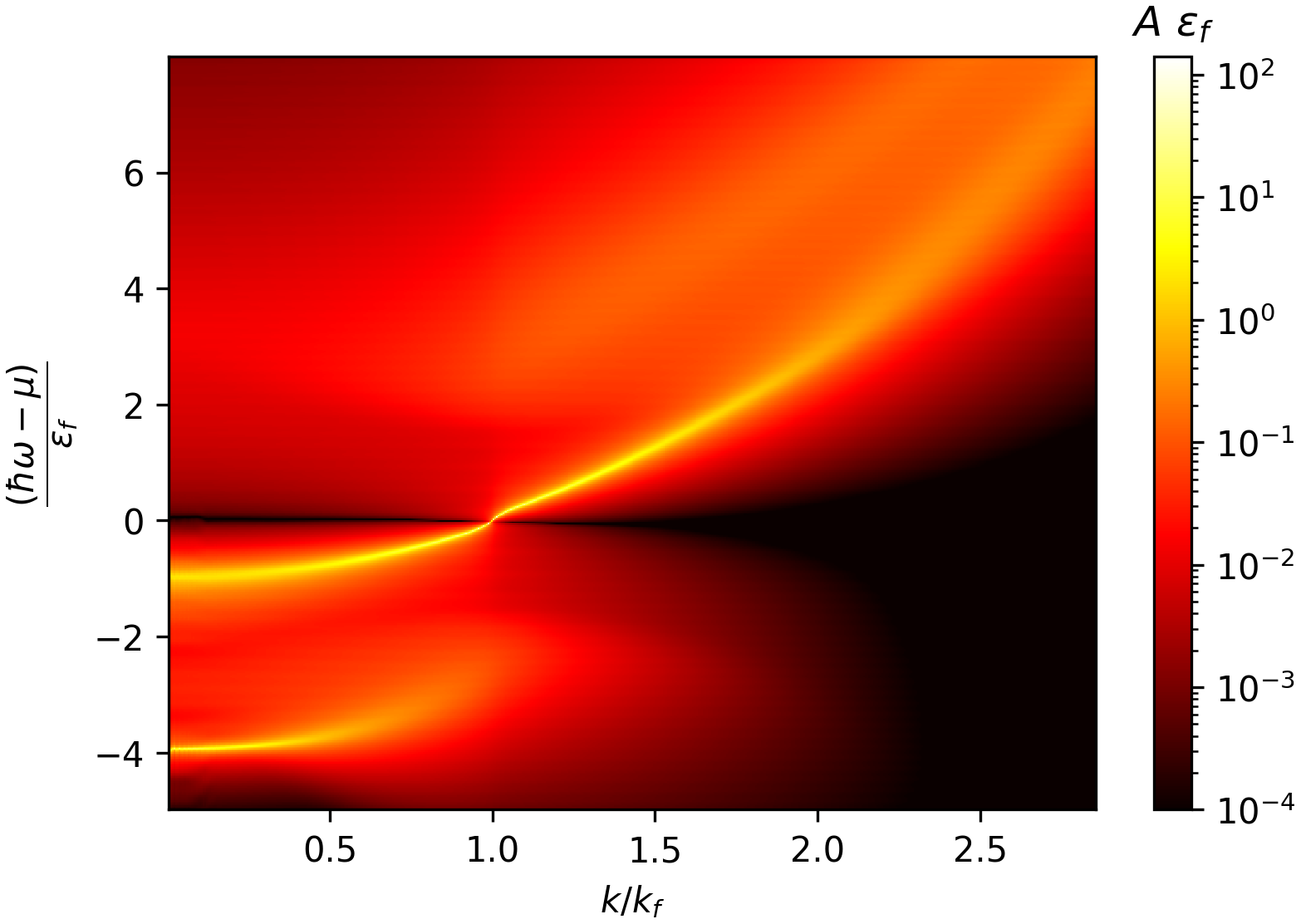}
    \caption{Spectral function of the HEG at $r_s=4$ from a $G_0W_0$ calculation. The Fermi energy is $\epsilon_f=\frac{\hbar^2 k_f^2}{2 m_e}$ with $k_f$ the Fermi momentum. $\mu=\epsilon_f+\text{Re}{\Sigma(k_f,\epsilon_f)}=\epsilon_f(1-0.0545)$ is the chemical potential. The scale of the color-map is logarithmic.}
    \label{fig:spectralFuncG0W0rs4}
\end{figure}

We start by considering the independent particle polarizability $P_0(q,\omega)$ computed at the $G_0$ level.
In Fig.~\ref{fig:polarizabilityPlotCalsVsTheory} we compare the imaginary part of $P_0$, calculated using Eq.~\eqref{eq:polarizabilityInt} and represented on SOP (fitted to $2^\text{nd}$ order Lorentzians with NNLS and then evaluated on a frequency grid, see Secs.~\ref{sec:method_transform} and~\ref{sec:validation}), with its analytic expression~\cite{fetter_quantum_2003} (note that this is the only analytic result we use as a check -- all others are evaluated numerically). The $\delta\to0^+$ broadening used in $G_0$ in order to converge the momentum integration does not sensibly affect the calculations.
It is worth noting that the use of $2^\text{nd}$ order Lorentzians with respect to simple Lorentzians eases this convergence, providing for the same $\delta$ and $k$-grid spacing better agreement with the analytic result at $\delta=0$ (thermodynamic limit, see Sec.~\ref{sec:method_representation}). 
From the plot comparison we can qualitatively infer that the SOP approach, together with its numerical implementation, is working effectively in computing and representing the dynamical polarizability across a range of different values of $q$. 

Next, we look at the self-energy numerical procedures by examining directly the $G_0W_0$ spectral function as shown in Fig.~\ref{fig:spectralFuncG0W0rs4}. This is obtained evaluating Eq.~\eqref{eq:sigmaIntegral}, representing the self-energy on SOP with $2^\text{nd}$ order Lorentzians, using the algorithmic inversion for the self-energy, and then evaluating the Green's function on a frequency grid. Focusing the attention on the lower satellite as well as on the  quasi-particle band, we can see that Fig.~\ref{fig:spectralFuncG0W0rs4} compares well with~\cite{caruso_gw_2016,pavlyukh_dynamically_2020} (note that, at variance with~\cite{pavlyukh_dynamically_2020}, we use a logarithmic scale to represent the intensity of the spectral function, in order to highlight its structure). The plasmaron peak~\cite{caruso_gw_2016} is very visible for small momenta where the quasi-particle band broadens, while the satellite band in the occupied-frequency range ($\omega<\mu$) is sharper. As $k$ approaches $k_f$, the plasmaron disappears and the quasi-particle band becomes more peaked. At $k=k_f$ the spectral function presents the typical metallic divergence along the quasi-particle band, and occupied and empty satellites are almost of the same weight, in agreement with Ref.~\cite{von_barth_self-consistent_1996}. 
For $k>k_f$ satellites coming from empty states ($\omega>\mu$) become dominant along with the quasi-particle band, and no structure resembling a plasmaron hole appears.

\subsection{Frequency integrated quantities and thermodynamics} \label{sec:results_HEG_thermo}
%

\begin{figure}
    \centering
    \includegraphics[width=\columnwidth]{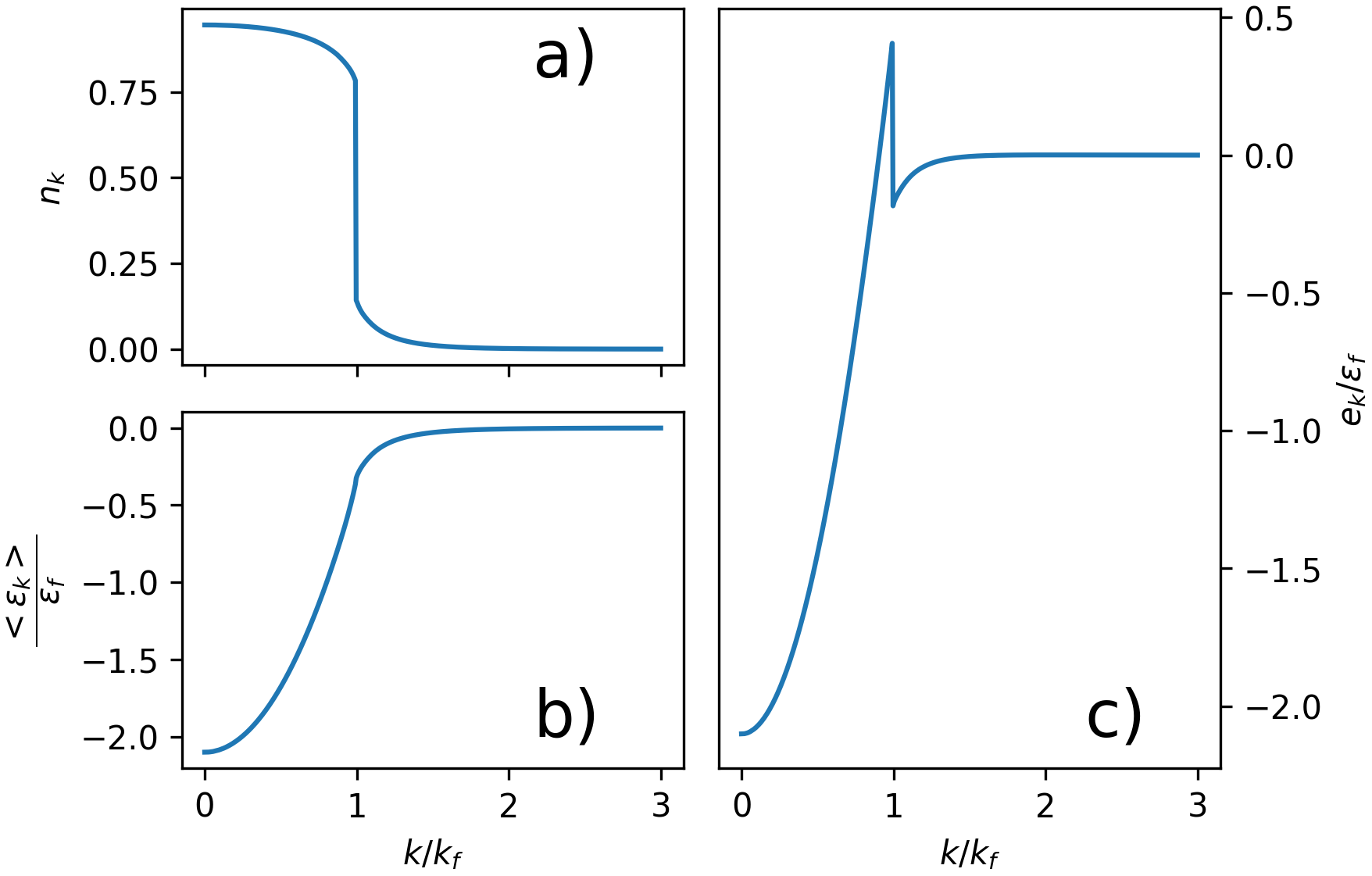}
    \caption{Selected frequency integrated quantities from a $G_0W_0$ calculation of the HEG at $r_s=4$. In panel a) the occupation number in arbitrary units, and in panel b) the occupied band $\expval{\epsilon_k}$ (see Sec.~\ref{sec:method_HEG_freqintegrated} for details), both as functions of the momentum $k$. In panel c) the Galitzki-Migdal total-energy resolved over $k$-contributions $e_k$, according to the rhs of Eq.~\eqref{eq:Galitski-Migdal} as function of the momentum $k$. $k_f$ and $\epsilon_f$ are the Fermi momentum and energy respectively.}
    \label{fig:frequencyIntegratedG0W0}.
\end{figure}

We now study convergence and stability of the total energies. Following the prescription of Sec.~\ref{sec:method_HEG}, we use the spectral function on SOP obtained in Sec.~\ref{sec:results_HEG_propagators}, Eq.~\eqref{eq:moments} to get analytically the occupation number $n_k$ and the occupied-band energy $\epsilon_k$ (see Sec.~\ref{sec:method_HEG_freqintegrated}), and finally numerically integrate the momenta of Eq.~\eqref{eq:Galitski-Migdal} to obtain the total energy. 
To perform the convergence study on the total energy, we follow the approach described in Sec.~\ref{sec:numerical_details} which consists in converging all parameters for the calculation separately.
Being the HEG a metal, the use of the algorithmic-inversion method to get a spectral function that obeys all sum rules (implied by the Dyson equation, see Sec.~\ref{sec:Dyson_SOP} for details), including the normalization condition for the spectral function, is crucial for obtaining  well-converged results. Indeed, the Luttinger discontinuity of $n_k$ makes the value of the total energy from the Galitzki-Migdal very sensitive to the converging parameters. 

In Fig.~\ref{fig:totEnecG0W0rs4} we show the convergence study for the correlation energy per particle (total energy minus Fock-exchange): the convergence value for $r_s=4$ is $0.0381 \pm 0.0003 \ \mathrm{Ha}$ in agreement with Refs.~\cite{garcia-gonzalez_self-consistent_2001} (with a difference of $0.0003$ Ha), where calculations were done along the imaginary axis. In panel a) of Fig.~\ref{fig:frequencyIntegratedG0W0} we plot $n_k$, and in panel b) $\expval{\epsilon_k}$ (as defined in Sec.~\ref{sec:method_HEG_freqintegrated}). The occupation number $n_k$ presents a sharp Luttinger discontinuity, which indicates that the broadening used in Eq.~\eqref{eq:sigmaIntegral} is well-controlled and does not spoil the quality of the results. In panel c) of Fig.~\ref{fig:frequencyIntegratedG0W0} we plot the total-energy resolved over $k$-contributions $e_k$ [rhs of Eq.~\eqref{eq:Galitski-Migdal}]. As previously mentioned, due to the presence of the Luttinger discontinuity, this function is sharp and thus difficult to integrate, at variance, e.g. with the RPA-Klein-energy functional, which is expected to be smoother~\cite{almbladh_variational_1999}.

\subsection{$G_0W_0$ for a broad range of HEG densities}
\label{sec:results_HEG_GoWo}

\begin{figure*}
    \centering
    \includegraphics{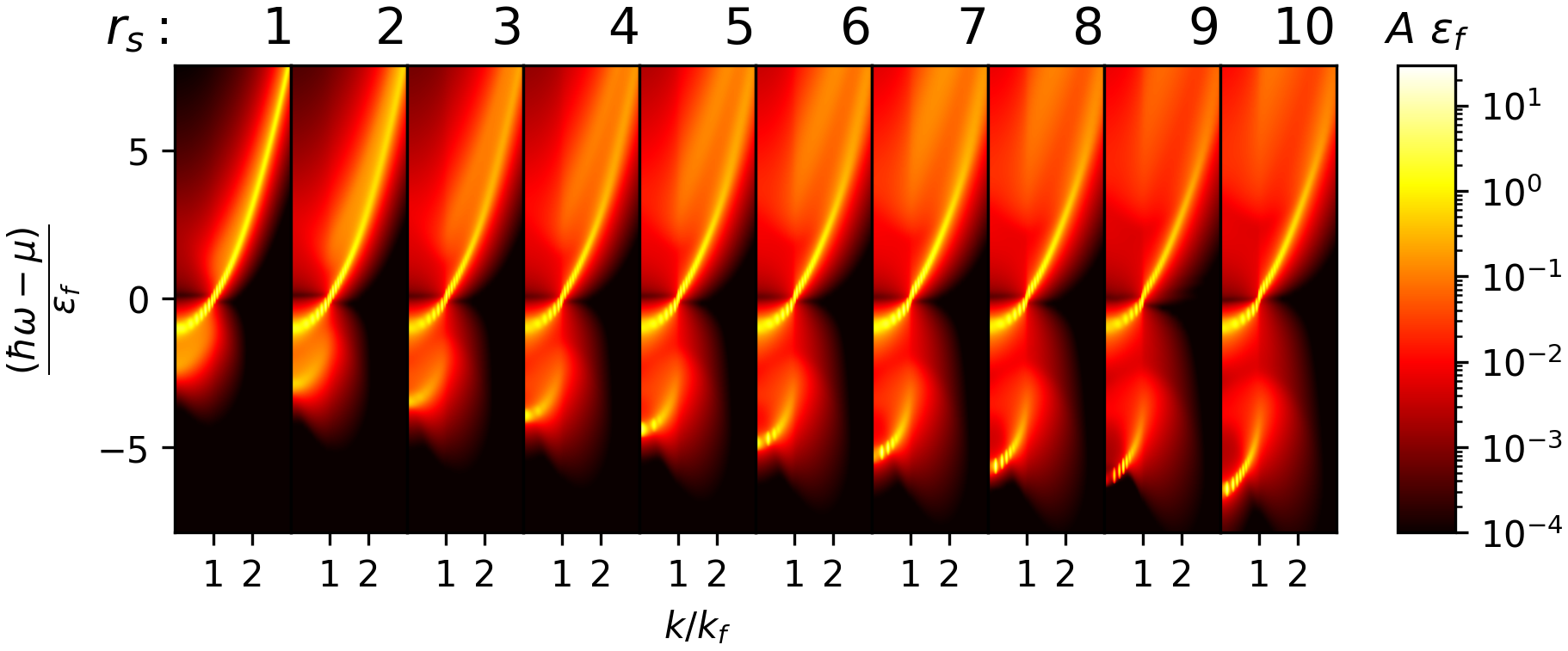}
    \caption{Spectral functions of the HEG at several densities. At the top $r_s$ specifies the density. The Fermi energy is $\epsilon_f=\frac{\hbar^2 k_f^2}{2 m_e}$ with $k_f$ the Fermi momentum. $\mu=\epsilon_f+\text{Re}\Sigma(k_f,\epsilon_f)$ is the chemical potential. The color map is logarithmic.}
    \label{fig:spectralFuncG0W0severalrs}
\end{figure*}

\begin{figure}
    \centering
    \includegraphics[width=\columnwidth]{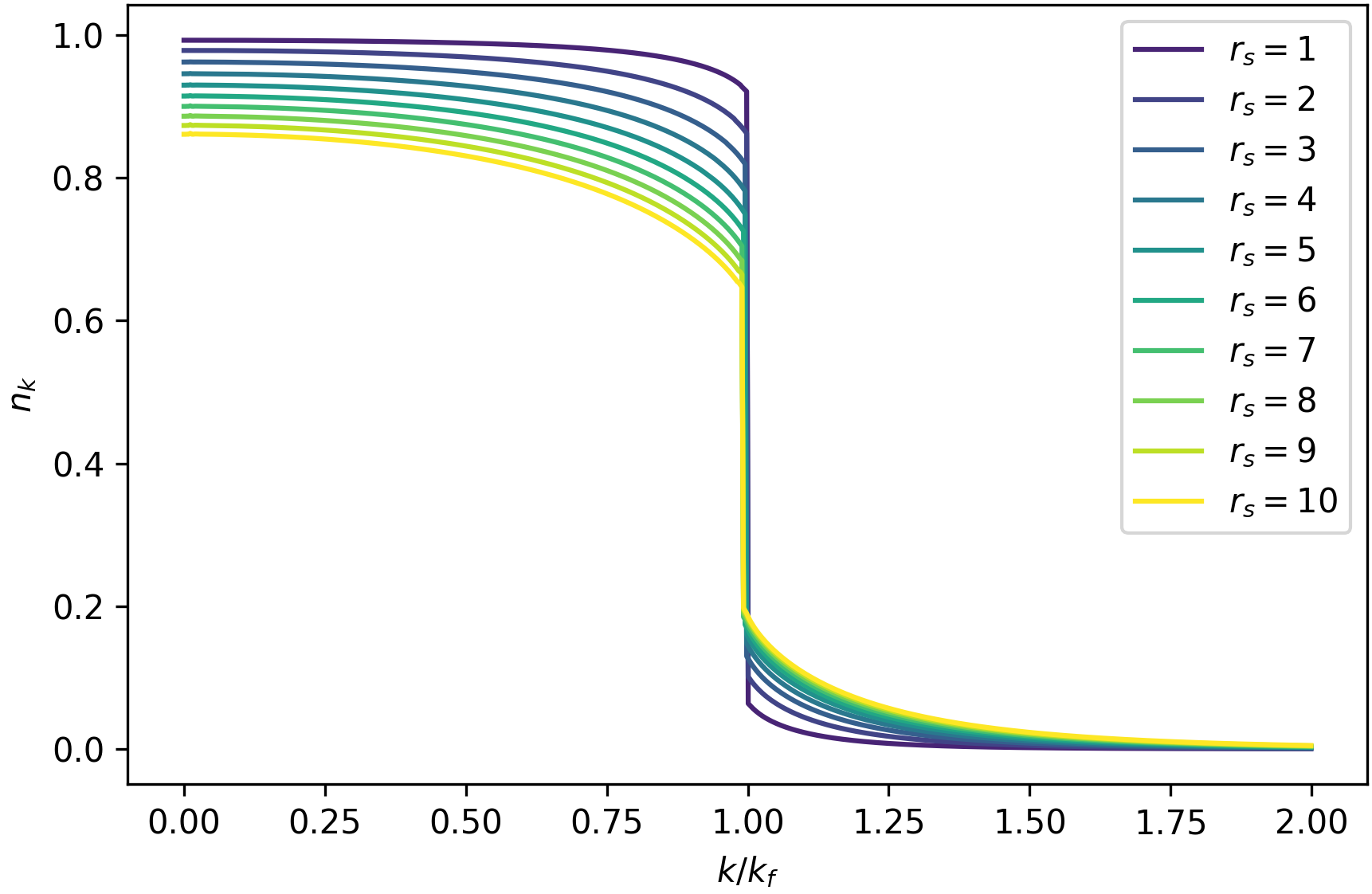}
    \caption{Occupation factor in arbitrary units for several densities ($r_s $ from $1$ to $10$). $k_f$ is the Fermi momentum.}
    \label{fig:occFactseveralrs}
\end{figure}

\begin{table}
    \begin{ruledtabular}
        \begin{tabular}{cccc}
         & \multicolumn{3}{c}{\bf $G_0W_0$ HEG Correlation Energies $|E_\text{corr}|$} \\[2pt]
        \hline
            $r_s$ 
            & {\bf This work} & {\bf Ref.}~\cite{holm_total_2000} & {\bf Ref.}~\cite{garcia-gonzalez_self-consistent_2001}\\
            \hline
                 $1$ &  $0.0749\ (\pm0.0015)$ &               $0.0722$ &                                 $0.0690$ \\
                 $2$ &  $0.0545\ (\pm0.0003)$ &               $0.0539$ &                                 $0.0530$ \\
                 $3$ &  $0.0451\ (\pm0.0008)$ &               $0.0448$ &                                        - \\
                 $4$ &  $0.0381\ (\pm0.0003)$ &               $0.0382$ &                                 $0.0378$ \\
                 $5$ &  $0.0333\ (\pm0.0002)$ &               $0.0355$ &                                 $0.0331$ \\
                 $6$ &  $0.0297\ (\pm0.0002)$ &                      - &                                        - \\
                 $7$ &  $0.0268\ (\pm0.0002)$ &                      - &                                        - \\
                 $8$ &  $0.0245\ (\pm0.0002)$ &                      - &                                        - \\
                 $9$ &  $0.0226\ (\pm0.0002)$ &                      - &                                        - \\
                $10$ &  $0.0210\ (\pm0.0002)$ &                      - &                                 $0.0207$ \\
    \end{tabular}
    \end{ruledtabular}
    \caption{Correlation energies as function of $r_s$ for the HEG at the $G_0W_0$ level. Energies are in Hartree units.}
    \label{tab:corelation_ene}
\end{table}{}

\begin{table}
    \begin{ruledtabular}
        \begin{tabular}{ccc}
            $\gamma$ & $\beta_1$ & $\beta_2$\\
            \midrule[0.3pt]
                $-0.1929$ & $1.1182$ & $0.4609$\\
            \midrule[0.3pt]
            & Covariance matrix of the fit & \\
            \midrule[0.3pt]
                $0.00022$ & $-0.00277$ & $-0.00011$ \\
                          & $0.03497$  & $0.00123$ \\
                          &            & $0.00014$ \\
    \end{tabular}
    \end{ruledtabular}
    \caption{Parameters of the correlation energy fit in Hartree, Eq.~\eqref{eq:perdew_zunger_fit} (same functional form as in~\cite{perdew_self-interaction_1981}),  using the data of Table~\ref{tab:corelation_ene}, and the covariance matrix of the fit. The fitted function is plotted in Fig.~\ref{fig:table_tot_ene_complete}.}
    \label{tab:fit_parameters_PZ}
\end{table}

\begin{figure}
    \centering
    \includegraphics[width=\columnwidth]{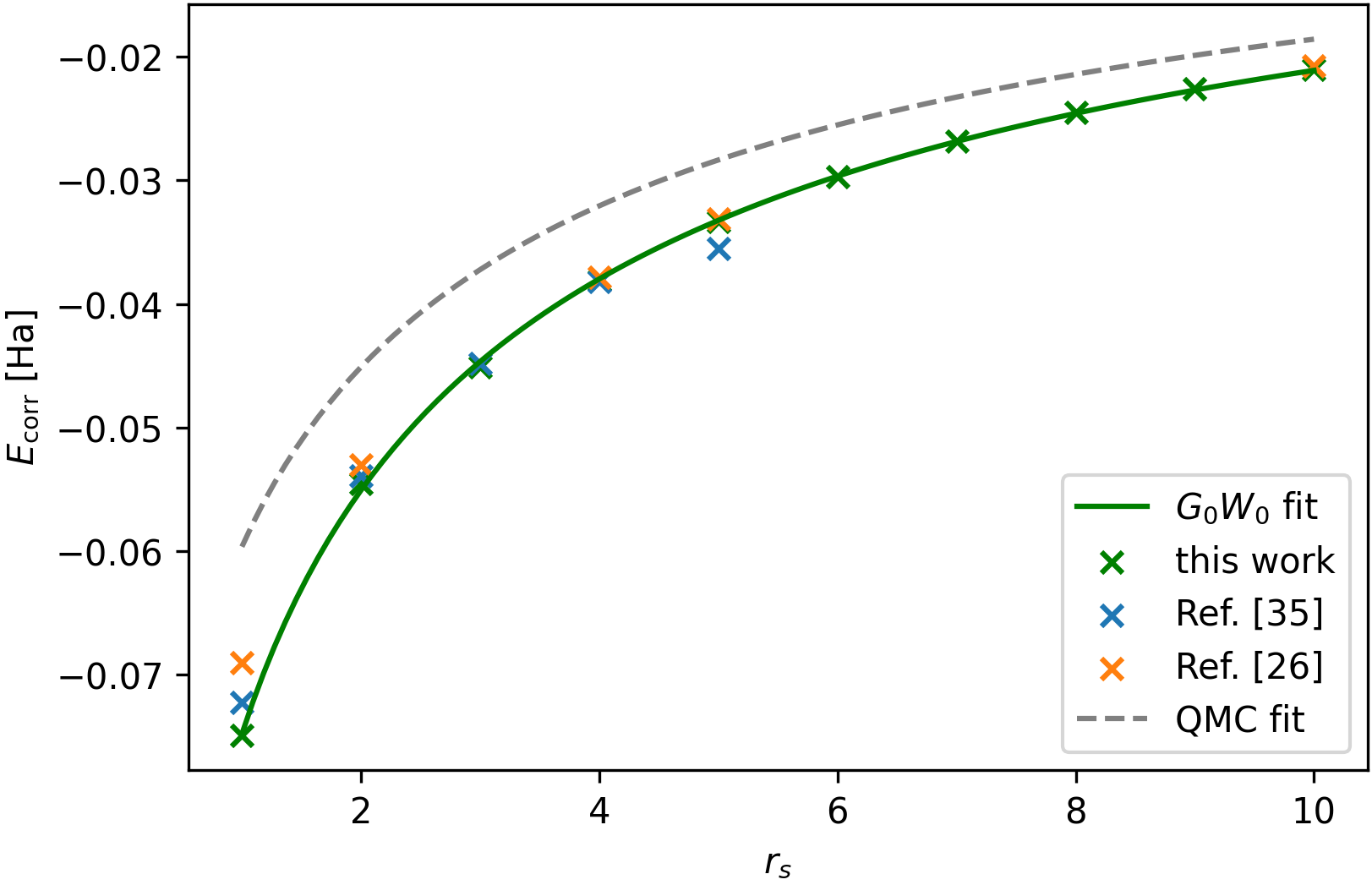}
    \caption{Correlation energy in Hartree units for several densities within the $G_0W_0$ approximation in the HEG. In green we show the results found in this work (see Sec.~\ref{sec:results_HEG_GoWo}) compared with those found in~\cite{holm_total_2000} in blue, and with~\cite{garcia-gonzalez_self-consistent_2001} in orange. In green we plot the correlation energy fit of Eq.~\eqref{eq:perdew_zunger_fit} (same functional form as in~\cite{perdew_self-interaction_1981}) on the present (green) data. For reference, in dashed grey we add also the Quantum Monte Carlo data obtained by Ceperley and Alder~\cite{ceperley_ground_1980} in the fit made by Perdew and Zunger~\cite{perdew_self-interaction_1981}.}
    \label{fig:table_tot_ene_complete}
\end{figure}

In this Section we report results for the HEG with $r_s$ ranging from $1$ to $10$ studied at the $G_0W_0$ level, following the same approach used for $r_s=4$.
In Fig.~\ref{fig:spectralFuncG0W0severalrs} we show the computed data for the spectral function obtained with the AIM-SOP approach. In the chosen units ($\epsilon_f$ for the energy and $k_f$ for the momentum) the spectral function for increasing $r_s$ shows an increase in the separation between the quasi-particle band and the satellite occupied and empty bands. Indeed, in these units $r_s$ controls the interaction strength ---see Eq.~(3.24) of Ref.~\cite{fetter_quantum_2003}--- with the limits of the non-interacting gas obtained for $r_s\to0$ and the strongly interacting gas corresponding to $r_s\to\infty$. Accordingly, the plasmaron peak of the satellite band at small momenta is weakened for smaller $r_s$. The same behaviours can be observed for the occupation factor of Fig.~\ref{fig:occFactseveralrs} for the different densities. For $r_s\to0$ the HEG approaches the non-interacting limit and the occupation number drops from $1$ to $0$ for increasing $k/k_f$. Going toward $r_s=10$ the jump becomes smaller, since the quasi-particle is reduced due to the more evident satellite bands, as it can be seen from Fig.~\ref{fig:spectralFuncG0W0severalrs}.

In Table~\ref{tab:corelation_ene} we report the corresponding total energies computed at the different densities, together with the available (to our knowledge) results in the literature. Since the calculations of Ref.~\cite{garcia-gonzalez_self-consistent_2001} were done on the imaginary axis, we shall consider those as the most accurate for the comparison. 
We refer to the \suppinfo\, for the convergence studies of the total energies for the different densities.
We find at $r_s=1$ the largest discrepancy ($0.0059$ Ha) with respect to the data of Ref.~\cite{garcia-gonzalez_self-consistent_2001}. This can be rationalized by noting, e.g., that $n_k$ is a steeper function, thereby enhancing the numerical issues of the Galitzki-Migdal expression discussed in Sec.~\ref{sec:method_HEG_freqintegrated}. To deepen the understanding of this numerical discrepancy, aside the convergence study of Fig.~\ref{fig:SI_totEnecG0W0rs1} provided in the \suppinfo, we performed an additional calculation increasing the refinement parameter $\Delta$ by $20\%$, aiming at increasing the accuracy in the integral grids, to target the steeper character of $r_s=1$. The result, $0.0736$ Ha against $0.0749$ Ha of Table~\ref{tab:corelation_ene}, is acceptable considering the error of $0.0015$ Ha of Table~\ref{tab:corelation_ene}. Most importantly we stress that at variance with Ref.~\cite{garcia-gonzalez_self-consistent_2001}, our procedure provides not only accurate frequency-integrated quantities (e.g. the total energy), but also precise spectral properties on the real axis (key quantities for spectroscopy).

In Fig.~\ref{fig:table_tot_ene_complete} we plot the correlation energy of Table~\ref{tab:corelation_ene} as a function of $r_s$, including the Perdez-Zunger (PZ) fit of the Quantum Monte Carlo (QMC) Ceperley Alder data as a reference~\cite{perdew_self-interaction_1981,ceperley_ground_1980}. We also exploit the same functional form of PZ to fit our data, providing in Table~\ref{tab:fit_parameters_PZ} $\gamma$, $\beta_1$, and $\beta_2$ for the fitting function for the correlation energy of the HEG (in Hartree):
\begin{equation}
    E_\mathrm{corr}(r_s)=\frac{\gamma}{1+\beta_1\sqrt{r_s}+\beta_2 r_s},
    \label{eq:perdew_zunger_fit}
\end{equation}
together with the covariance matrix of the fit. In Fig.~\ref{fig:table_tot_ene_complete} we plot the result of the fit as a green line. 

\section{Conclusions}
\label{sec:conclusion}
In this work we introduce the novel algorithmic-inversion method on sum over poles (AIM-SOP) to handle frequency-dependent quantities in dynamical theories.
Specializing to the case of many-body perturbation theory, we show that the AIM-SOP is able to provide a unified formalism for spectral and thermodynamic properties of an interacting-electron system.
Expanding all frequency-dependent quantities on SOP, we use AIM-SOP to solve exactly and at all frequencies Dyson-like equations, getting analytic frequency-dependent (spectral) and frequency-integrated (thermodynamic) properties. 
This is allowed by the mapping of the Dyson equation to an effective Hamiltonian of dimension controlled by the number of poles in the SOP of the self-energy (see Sec.~\ref{sec:Dyson_SOP}).
The transformation of frequency-dependent quantities into SOP is performed exploiting the representation of their spectral functions on different basis sets: aside from the standard choice of a basis of Lorentzians, we introduce $n$-th order generalized Lorentzian basis elements (see Sec.~\ref{sec:method_representation})
with improved decay properties. This allows for better numerical stability when transforming a propagator to SOP (see Sec.~\ref{sec:method_transform}), improved analytic properties for calculating the thermodynamic quantities (see Sec.~\ref{sec:method_analytical}), and an acceleration of convergence to the thermodynamic limit (zero broadening and infinite $k$-space sampling). Also, once the SOP representation of a propagator is known, we use the Cauchy residue theorem to calculate convolutions and (occupied) moments, accessing both spectral and thermodynamic quantities (see Sec.~\ref{sec:method_analytical}).

In order to have a working example of the AIM-SOP approach, we apply it to the paradigmatic case of many-body perturbation theory at the $G_0W_0$ level for the HEG at several densities ($r_s$ from $1$ to $10$). Using AIM-SOP, we are able to provide accurate spectra simultaneously with precise frequency-integrated quantities (e.g. occupation numbers and total energies). At the available densities,
we find very good agreement with Refs.~\cite{caruso_gw_2016,pavlyukh_dynamically_2020} for the spectral function. Moving to the total energy, we provide an in depth study of the stability and convergence of our results, finding quantitative agreement with Ref.~\cite{garcia-gonzalez_self-consistent_2001} for the available $r_s$, where calculations are performed on the imaginary axis.


Although in this article we study a homogeneous system as test case, the AIM-SOP approach aims to treat realistic non-homogeneous systems in the more general framework of dynamical embedding theories,
for a full-frequency representation of potentials and propagators, the flexibility for self-consistent calculations, and the exact solution of Dyson-like equations. 


\section{Acknowledgments}
\label{sec:Acknowledgments}

This work was supported by the Swiss National Science Foundation (SNSF) 
through grant No. 200021-179138 (T.C.) and its National Centre of 
Competence in Research MARVEL on ‘Computational Design and Discovery of 
Novel Materials’ (N.M.), and from the EU Commission for the MaX Centre 
of Excellence on ‘Materials Design at the eXascale’ under grant no. 
824143 (N.M., A.F.).
\appendix
\section{Sum-over-poles representation of an $n$-th order Lorentzian}
\label{sec:SOP_nLor}
In this Appendix we obtain the SOP representation of a Green's function from an $n$-th order Lorentzian spectral function.
Recalling Sec.~\ref{sec:method_representation}, the discrete time-ordered Hilbert transform (Eq.~\eqref{eq:TOHTdiscrete}) of a (not normalized) $n$-th order Lorentzian,
\begin{equation}
    \label{eq:TOHT_nlor_non_normalized}
    \int\frac{d\omega'}{\pi}\frac{1}{\omega-\omega'-i0^+\,  \text{sgn}(\epsilon_j)}\frac{\abs{\delta_j}^{2n-1}}{(\omega'-\epsilon_j)^{2n}+(\delta_j)^{2n}},
\end{equation}
induces a SOP representation for the Green's function, see Sec.~\ref{sec:method_representation}. The expression in Eq.~\ref{eq:TOHT_nlor_non_normalized}
can be computed using the residue theorem. Closing the contour in the upper/lower plane for $\omega \lessgtr\mu$, the poles of the integrand $\zeta_{j,m}=\epsilon_j+e^{i\frac{\pi}{2n}\left(1+2m\right)}\delta_j$ come only from the spectral function $A$. Using L'H\^{o}pital's rule, the residues of the integrand are reduced to
\begin{equation}
    R_{j,m} = -\frac{1}{2n\pi} \frac{e^{i\frac{\pi}{2n}\left(1+2m\right)}}{\omega-\zeta_{j,m}-i0^+\,  \text{sgn}(\epsilon_j)}.
\end{equation}
Thus, taking the limit for $\mathcal{C}$ on the real-axis, poles and residues of the SOP for $G$ are those in Eqs.~\eqref{eq:TOHT_nlor_basis_poles} and~\eqref{eq:TOHT_nlor_basis_res}. The normalization of the $n$-th order Lorentzian is given by summing $\alpha_{m}$ of Eq.~\eqref{eq:TOHT_nlor_basis_res} and using the geometric sum,
\begin{equation}
        N_n = -\frac{i}{n} \sum_{m=0}^{n-1} e^{i\frac{\pi}{2n}\left(1+2m\right)}
         = \frac{1}{n \sin{\left(\frac{\pi}{2n}\right)}}.
    \label{eq:sum_ak}
\end{equation}

\section{Moments of a propagator and occupied moments of its spectral function}
\label{sec:Moments_and_occupiedM}
In this Section we discuss the equality between the (regularized) moments of a propagator Eq.~\eqref{eq:moments} and the occupied moments of its spectral function.
For simplicity of notation we restrict to the case of a single $n$-th Lorentzian $\mathcal{L}_{\delta}^n$, as defined in Eq.~\eqref{eq:nlor_basis}, and focus on the $m=2(n-1)$ case [again here we suppose the integral in Eq.~\eqref{eq:moments} converges which is assured by $m=0$ and $m=1$, but must be stronger regularized for higher degrees]:
\begin{eqnarray}
\label{eq:occupiedMoments}
    E_{2(n-1)}[G] &=&
    \int_{-\infty}^{+\infty} \frac{d\omega}{2\pi i}
    \int d\omega'\,\frac{e^{i\omega0^+}\omega^{2(n-1)}}{\omega-\omega'}
    \mathcal{L}_{\delta_j}(\omega'-\epsilon_j) 
    \nonumber 
    \\
%
%
    &=&\int_{-\infty}^{\mu} d\omega'\, e^{i\omega'0^+} (\omega')^{2(n-1)} \mathcal{L}_{\delta_j}(\omega'-\epsilon_j)
    \nonumber \\
    &=&\int_{-\infty}^{\mu} d\omega\, \omega^{2(n-1)} A(\omega)
\end{eqnarray}
where $A(\omega)$ is the spectral function of $G$. To go from the second to the third line, we used the $1/\omega^{2n}$ decay of the $n$-th Lorentzian, and applied the dominated convergence theorem which allows for the $0^+$ limit to be performed inside the integral. The same derivation holds for lower degree moments.
For the higher order moments, $m>2(n-1)$ it is not possible to discard the $e^{i\omega0^+}$ factor in the integral, thus $E_{m>2(n-1)}[G]$ becomes complex.
The equality between $E_{m>2(n-1)}[G]$ [first and second line of Eq.~\eqref{eq:occupiedMoments}] and the occupied moments of $A$ [third line of~\eqref{eq:occupiedMoments}] is lost, with the integral for the occupied moments of $A$ diverging. 
The divergence happens because we cannot exchange the limit of the finite representation (controlled by $\delta_i$) and the lower bound $a\to-\infty$ of the integral $\int_{a}^{\mu} d\omega\, \omega^{2(n-1)} A(\omega)$. 
Numerically, this translates into performing the two limits in order, i.e. fixing the lower bound of the integral and controlling the integral stability for $\delta_i\to0$, then lower $a$ and again convergence the result for $\delta_i\to0$, and repeat until both convergences are achieved. 
In this continuous limit for the representation of $A$ the integral for the occupied moments of $A$ [third line of Eq.~\eqref{eq:occupiedMoments}] coincides with $E_{m}[G]$ [last line of Eq.~\eqref{eq:moments}], thus the equality between the two is recovered.

\section{Exploiting parity of the RPA-polarizability integral}
\label{sec:parity_polariz_integral}
%
In this Appendix we show how it is possible to exploit the parity of the polarizability $P(q,\omega)$ at fixed momentum $\mathbf{q}$.
As explained in Sec.~\ref{sec:method_analytical}, the SOP approach allows to compute analytically the convolution of Eq.~\eqref{eq:polarizabilityIntStandard}. Using Eq.~\eqref{eq:convolutionSOP} in Eq.~\eqref{eq:polarizabilityIntStandard}, the polarizability may be rewritten as
\begin{multline}
    P(q,\omega)=2\int \frac{d\mathbf{k}}{(2\pi)^3} \times \\
    \Bigg[ \sum_{\substack{i,j \\ \Im{z_i(\abs{\mathbf{k+q}})}<0  \\ \Im{{z}_j(k)}>0 }} \frac{A_i(\abs{\mathbf{k+q}}) {A}_j(k)}{\omega+{z}_j(k)-z_i(\abs{\mathbf{k+q}})}\\
    - \sum_{\substack{i,j \\ \Im{z_i(\abs{\mathbf{k+q}})}>0 \\ \Im{{z}_j(k)}<0 }} \frac{A_i(\abs{\mathbf{k+q}}) {A}_j(k)}{\omega +{z}_j(k)-z_i(\abs{\mathbf{k+q}})} \Bigg],
    \label{eq:polarizabilityIntConvolved}
\end{multline}
where we did not yet restrict to the $G_0$ case in which only one pole is present. Calling $I(\abs{\mathbf{k+q}}_\text{unocc},k_\text{occ},\omega)$ the first term in the rhs ($k_\text{occ}$ labels the occupied states with momentum $k$, while $\abs{\mathbf{k+q}}_\text{unocc}$ refers to empty states), and setting $\mathbf{k+q} \to -\mathbf{k}$ in the second term,
\begin{multline}
    P(q,\omega)=2\int \frac{d\mathbf{k}}{(2\pi)^3} 
    \, \Big[ I(\abs{\mathbf{k+q}}_\text{unocc},k_\text{occ},\omega)\\
    +I(\abs{\mathbf{k+q}}_\text{unocc},k_\text{occ},-\omega) \Big]
    \label{eq:polarizabilityIntConvolvedIocc},
\end{multline}
it is possible limit the calculation to the first term. For the case of $G=G_0$ of Sec.~\ref{sec:method_HEG_propagators}, the occupied states at momentum $k$ are all within the Fermi sphere, and thus we can limit the momentum integration to the sphere of radius $k_f$, i.e. $k \le k_f$ in Eqs.~\eqref{eq:polarizabilityIntStandard} and~\eqref{eq:polarizabilityInt}.


%
%
\bibliographystyle{my_aip}
\bibliography{references.bib,references_extra.bib}
\makeatletter\@input{supplemental_material_aux_manual.tex}\makeatother

\end{document}